\documentclass{jpp}
\usepackage{graphicx}

\usepackage[utf8]{inputenc}
\usepackage[T1]{fontenc}
\usepackage{amsmath}
\usepackage{placeins}
\usepackage[normalem]{ulem}
\usepackage{url}

\newcommand{\stkout}[1]{\ifmmode\text{\sout{\ensuremath{#1}}}\else\sout{#1}\fi}

\shorttitle{A New Optimized Quasihelically Symmetric Stellarator}
\shortauthor{A. Bader et al}

\title{A New Optimized Quasihelically Symmetric Stellarator}

\author{A. Bader\aff{1}
 \corresp{\email{abader@engr.wisc.edu}},  
 B.J.~Faber\aff{1}, J.C. Schmitt \aff{2}, D.T.~Anderson\aff{1}, M.~Drevlak\aff{3}, J.M.~Duff\aff{1}, H.~Frerichs\aff{1}, C.C.~Hegna\aff{1}, T.G.~Kruger \aff{1}, M.~Landreman\aff{4} I.J.~McKinney\aff{1}, L.~Singh\aff{1}, J.M.~Schroeder\aff{1},  P.W.~Terry\aff{1}, A.S.~Ware\aff{5}}

\affiliation{\aff{1}University of Wisconsin-Madison, Wisconsin, USA
\aff{2} Auburn University, Alabama, USA
\aff{3} IPP-Greifswald, Germany
\aff{4} University of Maryland-College Park, Maryland, USA
\aff{5} University of Montana, Montana, USA}

\begin{document}

\maketitle

\begin{abstract}
A new optimized quasihelically symmetric configuration is described that has the desirable properties of improved energetic particle confinement, reduced turbulent transport by 3D shaping, and non-resonant divertor capabilities. The configuration presented in this paper is explicitly optimized for quasihelical symmetry, energetic particle confinement, neoclassical confinement, and stability near the axis. Post optimization, the configuration was evaluated for its performance with regard to energetic particle transport, ideal magnetohydrodynamic (MHD) stability at various values of plasma pressure, and ion temperature gradient instability induced turbulent transport. The effect of discrete coils on various confinement figures of merit, including energetic particle confinement, are determined by generating single-filament coils for the configuration. Preliminary divertor analysis shows that coils can be created that do not interfere with expansion of the vessel volume near the regions of outgoing heat flux, thus demonstrating the possibility of operating a non-resonant divertor.
\end{abstract}

\section{Introduction}
    This paper discusses results from optimizations to produce quasihelically symmetric (QHS) equilibria that  simultaneously demonstrate multiple desirable properties for advanced stellarators.  These properties include excellent neoclassical and energetic particle confinement, a reduction in turbulent transport, and a functional non-resonant divertor \citep{bader2017hsx, boozer2018simulation}. The baseline targets for the configuration also avoided low order rational surfaces, and included a vacuum magnetic well to avoid interchange instabilities. In this paper, we present a configuration that includes both the desired confinement and global macroscopic properties.
    
    Optimization studies of stellarators have a long and rich history\citep{grieger1992physics}.  The two constructed optimized stellarator experiments to date are the quasihelically (QH) symmetric Helically Symmetric eXperiment (HSX) \citep{anderson1995helically} and the quasi-omnigenous Wendelstein 7-X (W7-X) \citep{beidler1990physics}. Examples of quasiaxisymmetric (QA) stellarator concepts are the National Compact Stellarator eXperiment (NCSX) \citep{zarnstorff2001physics} and the Chinese First Quasi-axisymmetric Stellarator \citep{shimizu2018configuration}. Optimized stellarator configurations have also been the focus of reactor concepts. These include the ARIES-CS project \citep{ku2008physics} adapted from the NCSX design, the HELIAS reactor concept \citep{beidler2001helias} adapted from W7-X, and the Stellarator Pilot Plant Study \citep{miller1996stellarator} a QH configuration adapted from HSX.
    
    Improvements in optimization tools have led to new configuration designs. These advances come in several forms including the identification of new QA configurations \citep{henneberg2019properties}. Advances in theoretical understanding have produced mechanisms for optimization in the areas of turbulent transport \citep{mynick2010optimizing, xanthopoulos2014controlling,hegna2018theory}, energetic particle transport \citep{bader2019stellarator, henneberg2019improving}, and non-resonant divertors \citep{bader2018minimum}. Also, construction of optimized equilibria from first principles has been demonstrated for quasisymmetric stellarators \citep{landreman2018direct, landreman2019direct} and quasi-omnigenous stellarators \citep{plunk2019direct}.

    The structure of the paper is as follows.  Section \ref{sec:opt} provides a more detailed description of the optimization process for stellarators. Section \ref{sec:coils} describes the generation of coils for the device. In section \ref{sec:eval}, detailed properties of the device are examined. These include, in order, energetic particle transport, turbulent transport, MHD stability, and divertor construction. A conclusion and a description of future work is given in section \ref{sec:conc}. Additionally Appendix \ref{appA} provides a description of an optimized configuration evaluated for a midscale device.

\section{Optimization}
\label{sec:opt}

Typically, stellarators are optimized by representing the plasma boundary in a two-dimensional Fourier series and perturbing that boundary in an optimization scheme.  The boundaries for stellarator symmetric equilibria are given as,
\begin{equation}
\label{eq:boundary}
R = \sum_{m,n} R_{m,n}\mathrm{cos}\left( m\theta - n\phi\right);\; Z = \sum_{m,n} Z_{m,n}\mathrm{sin}\left( m\theta - n\phi\right)    
\end{equation}
Here $(R,Z,\phi)$ represent a cylindrical coordinate system, $\theta$ is a poloidal-like variable, and $R_{m,n}$, $Z_{m,n}$ are the Fourier coefficients for the $m$th poloidal and $n$th toroidal mode numbers.  This representation enforces stellarator symmetry, as is commonly used in stellarator design.

Using the boundary defined in \ref{eq:boundary} and profiles for the plasma pressure and current, the equilibrium can be solved for at all points inside the boundary. Equilibrium solutions in this paper are calculated using the Variational Moments Equilibrium Code (VMEC) \citep{Hirshman_PoF_1983}. The full equilibrium can then be evaluated for various properties of interest and the overall performance of the configuration can thus be determined.

Optimization of the equilibrium is performed by the ROSE (Rose Optimizes Stellarator Equilibria) code \citep{drevlak2018optimisation}. ROSE evaluates boundaries by first computing a VMEC equilibrium solution and then applying user defined metrics with appropriate weights. A target function for the equilibrium is computed as
\begin{equation}
    F\left(\mathsfbi{R}, \mathsfbi{Z}\right) = \sum_i \left(f_i\left(\mathsfbi{R}, \mathsfbi{Z}\right) - f_i^\mathrm{target} \right)^2 w_i\left(f_i\right) \sigma_i\left(f_i\right)
\end{equation}
Here, $\mathsfbi{R}$ and $\mathsfbi{Z}$ represent the arrays of the $R_{m,n}$ and $Z_{m,n}$ coefficients that define the boundary. The summation is over the different penalty functions, the $f_i$s, chosen by the user, each of which has a weight, $w_i$, a target, $f_i^\mathrm{target}$ and a function $\sigma_i$ which determines whether the penalty function is a target or a constraint.  A target function tries to minimize $f_i - f_i^\mathrm{target}$ while a constraint only requires either $f_i < f_i^\mathrm{target}$ or $f_i > f_i^\mathrm{target}$. 

The variation of the boundary coefficients, $\mathsfbi{R}$ and $\mathsfbi{Z}$ is performed through an optimization algorithm. For the resulting configuration shown here, Brent's algorithm is used \citep{brent2013algorithms}. Additional details of this optimization technique can be found in \citep{drevlak2018optimisation, henneberg2019properties} and results for similar optimizations of quasisymmetric stellarators can be found in \citep{ku2008physics}.

The optimization for this new quasihelically symmetric configuration is an extension of a scheme that focused on improved energetic particle transport \citep{bader2019stellarator}. For the new optimization, a four field period device is chosen. The six metrics included in this optimization are the following:

\begin{itemize}
    \item The deviation from quasisymmetry, where the quantity to be minimized is defined as the energy in the non-symmetric modes normalized to some reference field, here the field on axis. In more detail, the configuration is converted into Boozer coordinates \citep{boozer1981coordinates} and the magnitude of the magnetic field strength is represented in a discrete Fourier series with coefficients $B_{m,n}$. Then the quasisymmetry deviation for a four field period device is calculated as,
    \begin{equation}
    \label{eq:qh}   
        P_{QH} = \left( \sum_{n/m \neq 4} B_{m,n}^2 \right) / B_{0,0}^2
    \end{equation}
    where $B_{0,0}$ is the field on axis. 
    
    \item The $\Gamma_c$ metric is a proxy for energetic particle confinement \citep{nemov2008poloidal}. In brief, this metric seeks to align contours of the second adiabatic invariant, $J_\parallel$ with flux surfaces. Its viability in producing configurations with excellent energetic ion confinement was demonstrated in \citep{bader2019stellarator}.
    
    \item A magnetic well \citep{greene1997brief}, where present, provides stability against interchange modes. Because finite $\beta$ effects in quasihelical symmetry tend to deepen magnetic wells, it suffices to have a vacuum magnetic well of any strength. In this optimization, the vacuum well was required to exist at the magnetic axis, but was not optimized further beyond that.
    
    \item The rotational transform profile was chosen to avoid low order rational surfaces, and to be sufficiently high above the $\stkout{\iota}$ = 1 surface to avoid a resonance when nominal amounts of plasma current are added.
    
    \item The aspect ratio of the configuration was fixed at the starting aspect ratio of 6.7. No deviations to the aspect ratio were allowed beyond that value.
    
    \item The neoclassical transport metric, $\epsilon_\mathrm{eff}$ \citep{nemov1999evaluation} is required to remain below some nominal value, here 0.01.  In practice, adequate quasisymmetry enforces this constraint automatically. In addition, proxies that attempt to align the maxima and minima values of the magnetic field strength on a field line were also employed in the optimizations. These proxies compute the contour of maximum (or minimum) magnetic field on a surface and seek to minimize the variance of the magnetic field strength on that contour.
    
    \item Although not explicitly optimized for, it was desired that the configuration has lower turbulent transport for the same profile values relative to that in HSX. Explicit optimization for turbulent transport is a subject of future work.

\end{itemize}
\subsection{Configuration characteristics}
\begin{figure}
  \centering
  \includegraphics[width=\textwidth]{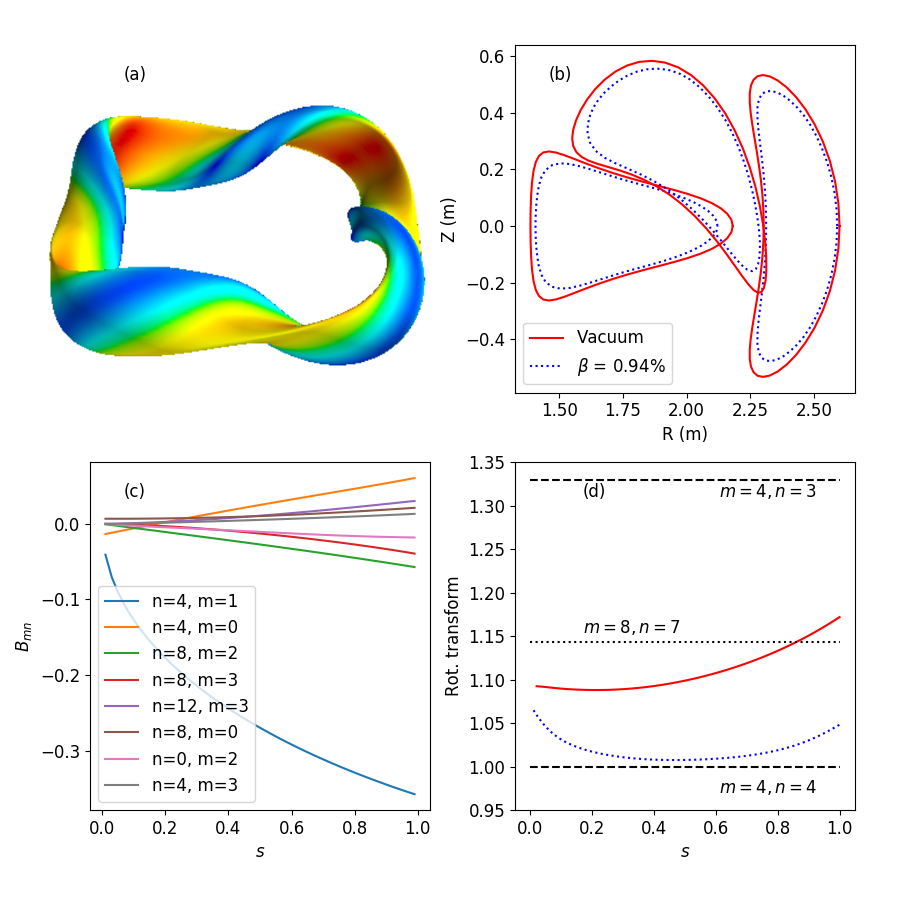}
  \caption{Contours of the magnetic field strength on the boundary are plotted in top left, (a). The top right plot, (b), shows boundary surfaces at toroidal cuts of toroidal angles 0, $\pi/8$ and $\pi/4$ for the vacuum configuration (red) and a configuration with 0.94\% $\beta$ (blue dot). The bottom left plot (c) shows the vacuum Boozer spectrum with the strengths of the 8 most dominant modes as a function of normalized toroidal flux, $s$. The bottom right plot, (d), shows the vacuum rotational transform profile (red) and the rotational transform profile at 0.94\% $\beta$ (blue dot). In (d) important rational surfaces are plotted with dashed and dotted black lines.}
\label{fig:wv4}
\end{figure}

A new optimized configuration, hereby termed the WISTELL-A configuration, has been generated through the optimization procedures described above. Figure \ref{fig:wv4} shows some of the properties of the configuration. Contours of the vacuum magnetic field strength on the boundary are shown in figure \ref{fig:wv4}a. The helical nature of the magnetic field strength, a feature of quasihelical symmetry, is clearly seen. Figure \ref{fig:wv4}b shows toroidal cuts of the boundary surface at toroidal angles 0, $\pi/8$ and $\pi/4$, colloquially referred to as the "bean", "teardrop" and "triangle" surfaces respectively.  Shown are both the vacuum boundary surfaces (red) and the surfaces with normalized plasma pressure $\beta = 0.94\%$. The surfaces with finite pressure are generated by a free boundary solution using coils and self consistent bootstrap current. The procedure will be described in section \ref{sec:mhd}. The finite pressure boundary is somewhat smaller than the vacuum boundary.  Figure \ref{fig:wv4}c shows the Boozer spectrum for the vacuum equilibrium where the $n=0, m=0$ mode has been suppressed. The $n=4, m=1$ mode is dominant, which is expected for a four-period quasihelically symmetric equilibrium. The largest nonsymmetric modes are the mirror mode at $n=4, m=0$ and the $n=8, m=3$ mode. Note that the $n=8, m=2$ mode is also symmetric. Figure \ref{fig:wv4}d shows the rotational transform profile both for vacuum at $\beta = 0.94$\%. The dashed black lines represent the major low order rational surfaces that should be avoided, these are the $\stkout{\iota}$ = 1.0 and $\stkout{\iota}$ = 4/3 surfaces.  The configuration passes through the $\stkout{\iota}$ = 8/7 surface at the edge in the vacuum configuration. This vacuum surface can possibly be used to test island divertor features, which will be discussed in section \ref{sec:div}. The blue dotted line in figure \ref{fig:wv4}d represents the rotational transform profile at $\beta = 0.94$\%. As can be seen, the minimum of the rotational transform profile is just above the $\stkout{\iota} = 1$ surface. 

\begin{figure}
  \centering
  \includegraphics[width=\textwidth]{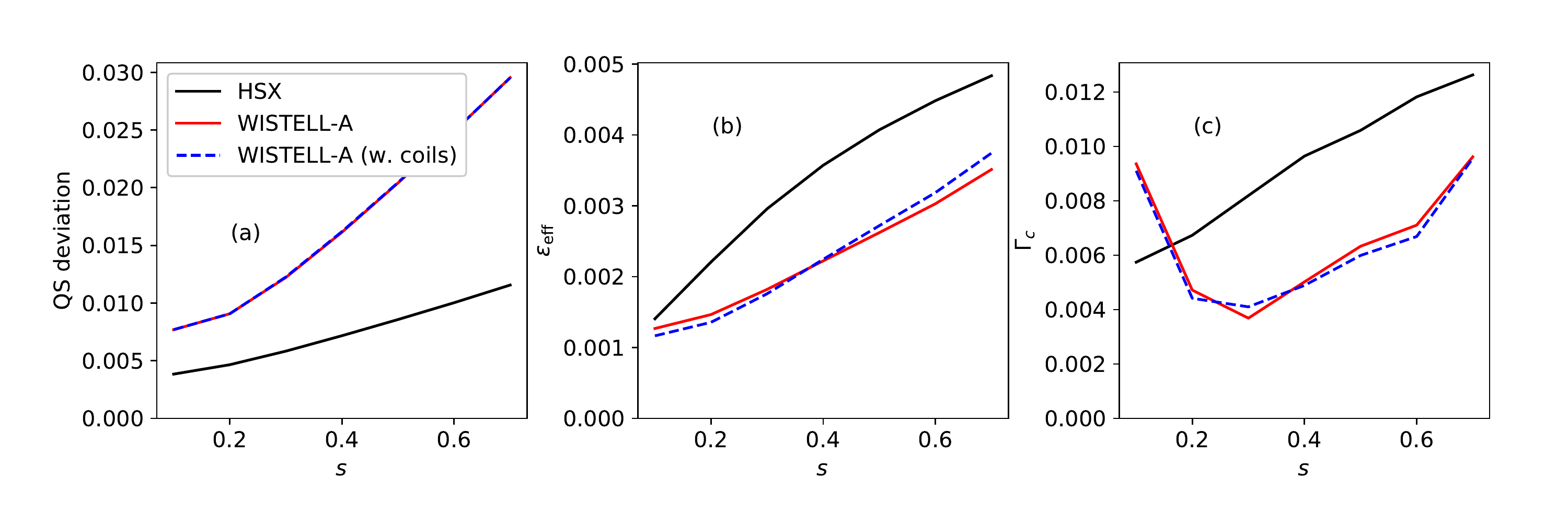}
  \caption{Values of the quasisymmetry deviation (left, a), $\epsilon_\mathrm{eff}$ (middle, b) and $\Gamma_c$ (right, c) are plotted as a function of normalized toroidal flux, $s$ for three configurations, the HSX configuration (black solid), the WISTELL-A configuration (red solid) and the WISTELL-A configuration as produced by coils (blue dashed).}
\label{fig:wv3}
\end{figure}

In addition, we note some derived features of the configuration. In figure \ref{fig:wv3}, we show the neoclassical transport, as quantified by the $\epsilon_{\mathrm{eff}}$ metric, the quasisymmetry deviation, as described in equation \ref{eq:qh}, and the $\Gamma_c$ metric for vacuum configurations. In order to provide a baseline for comparison, we include the same quantities calculated for the HSX equilibrium (black). HSX has better values obtained for the quasisymmetry metric, but slightly worse values for $\epsilon_{\mathrm{eff}}$ and $\Gamma_c$ over the majority of the minor radius. In addition to the optimized vacuum configuration, we also show the quantities for the vacuum fields produced by filamentary coils. The results here show that the configuration produced with coils does a very good job at reproducing the important qualities of the equilibrium. 

\section{Coil construction}
\label{sec:coils}

\begin{figure}
  \centering
  \includegraphics[width=\textwidth]{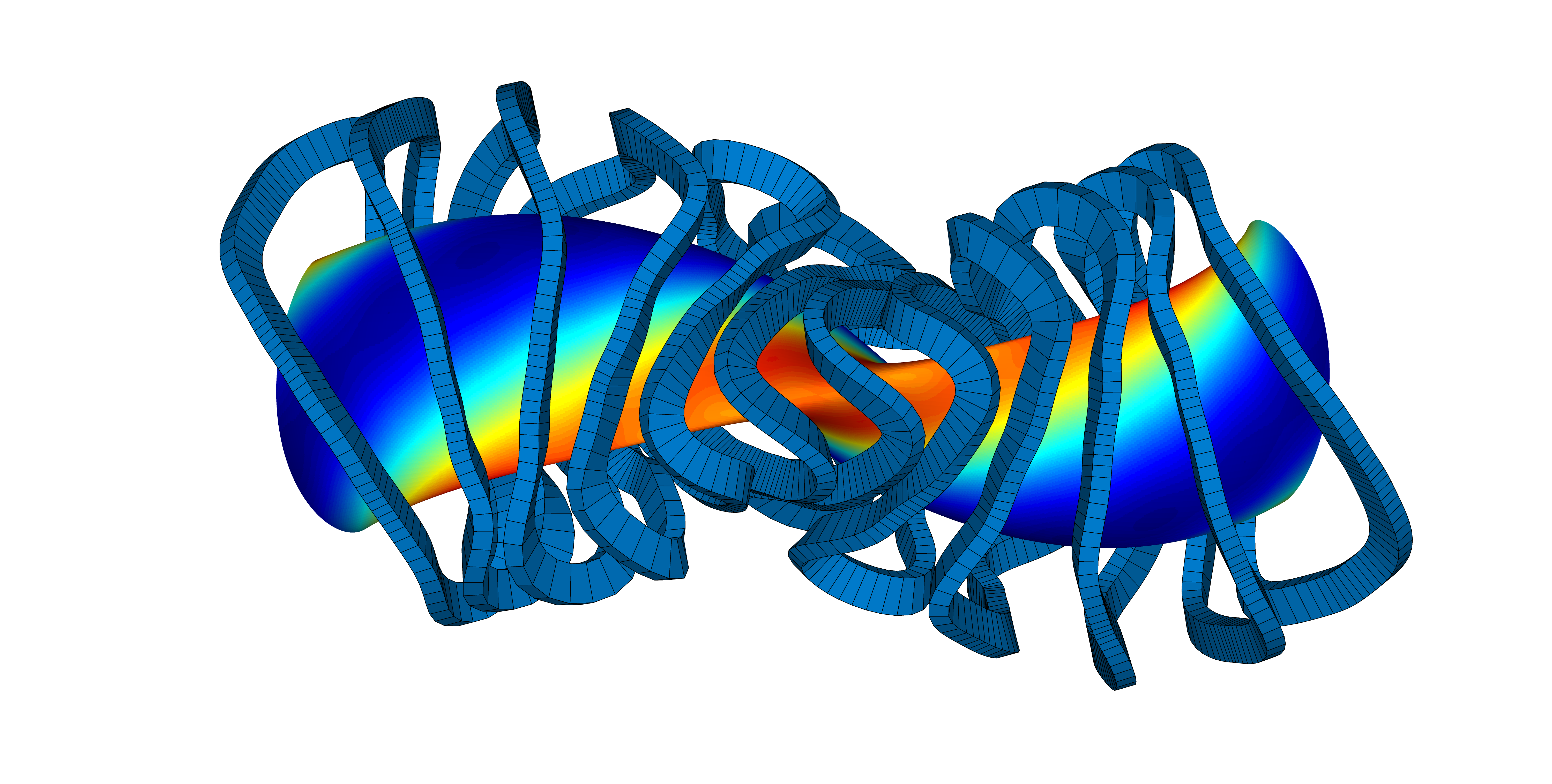}
  \caption{A representation of coils for the WISTELL-A configuration. Internal to the coils is a representation of the magnetic field strength on the boundary as produced by the coils.}
\label{fig:atencoils}
\end{figure}

Coils to reproduce the vacuum boundary were produced by the FOCUS code \citep{zhu2018designing} using an initial coil set generated by the REGCOIL code \citep{landreman2017improved}. The FOCUS code targets the average normal field on the magnetic boundary (a quantity to minimize) and the minimal radius of curvature for the coils (a quantity to maximize). A representation of the coils are given in figure \ref{fig:atencoils} along with the magnetic field magnitude on the boundary.

 Of particular importance is the fact that the FOCUS code does not require a specified coil winding surface, and is therefore able to move the coils further from the plasma in regions where the reliability of reproducing the configuration is less sensitive to the coil position. The ability to optimize without being confined to a winding surface is a new capability within FOCUS that was not available for the coil sets designed for HSX and W7-X. Previous results showed that areas of low sensitivity can be calculated for stellarator equilibria using shape gradients \citep{landreman2018computing}. Fortuitously, these regions of low sensitivity are also the areas were divertor heat fluxes tend to exit the plasma \citep{bader2017hsx}, allowing for the construction of a non-resonant divertor as described in section \ref{sec:div}.

\section{Performance evaluation}
\label{sec:eval}

The performance of the configuration is evaluated in several topical areas. These include confinement of energetic particles evaluated by Monte Carlo analysis, turbulent transport evaluated by non-linear \textsc{GENE} simulations \citep{jenko2000gene}, stability at finite pressure evaluated by COBRAVMEC \citep{sanchez2000cobra}, and an initial attempt at edge transport and divertor behavior evaluated by EMC3-EIRENE \citep{feng20043d}.

\subsection{Energetic particles}
\label{subsec:ep}

Energetic particle optimization was obtained in these configurations by targeting the $\Gamma_c$ metric that seeks to align contours of the second adiabatic invariant, $J_\parallel$ with flux surfaces. The calculation for $\Gamma_c$ is given in 
\citet[Eq. 61]{nemov2008poloidal} as
\begin{equation}
  \label{eq:biggammac}
  \Gamma_c = \frac{\pi}{\sqrt{8}} \lim_{L_s \rightarrow \infty} \left( \int_0^{L_s} \frac{ds}{B} \right)^{-1}
  \left[ \int_1^{B_{max}/B_{min}} db' \sum_{\mathrm{well}_j} \gamma_{cj}^2 \frac{v \tau_{b,j}}{4 B_{min} {b'}^2} \right].
\end{equation}
where the electric field contribution is ignored and the arbitrary reference field $B_0 = B_{min}$. The quantity $\gamma_c$ is
\begin{equation}
\label{eq:gammac}
  \gamma_c = \frac{2}{\pi} \mathrm{arctan} \left( \frac{v_r}{v_\theta} \right).
\end{equation}
Here, $v_r$ is the bounce averaged radial drift, $v_\theta$ is the bounce
averaged poloidal drift. The ratio $v_r/v_\theta$ is the key quantity to minimize. A method to calculate $v_r/v_\theta$ from geometrical quantities of the magnetic field line is described in \citet[Eq. 51]{nemov2008poloidal}. The summation in eq. \ref{eq:biggammac} is taken over all the wells for a suitably long field-line. In our case between 60 and 100 toroidal transits were used. The calculation considers trapping wells encountered by all possible trapped-particle pitch angles, with $b'$
representing a normalized value of the reflecting field.  The bounce
time for a particle in a specific magnetic well is given by
$\tau_{b,j}$. The parameters $B_{max}$ and $B_{min}$ are the maximum and minimum magnetic field strength on the flux surface. 

Previously, equilibria in ROSE were optimized by simultaneously minimizing $\Gamma_c$ and the quasihelical symmetry deviation.  This resulted in configurations with very low collisionless particle losses  \citep{bader2019stellarator}.
However, calculations to assess energetic confinement used ideal, fixed boundary equilibria and did not include the effects of coils. For the following evaluation of our new configuration,  energetic particle transport is calculated using the magnetic field structure produced from the filamentary coils presented in Sec.~\ref{sec:coils}.

To evaluate energetic particle transport, the transport of fusion-born alpha particles are examined in a configuration scaled to the ARIES-CS volume (450 m$^3$) and on-axis field strength (5.7 T). We choose a flux surface and distribute the alpha particles on the flux surface such that they properly resemble a distribution of alpha particles. The evaluation is done with the \texttt{ANTS} code \citep{drevlak2014fast}, which constructs a magnetic field grid in cylindrical coordinates. Therefore, \texttt{ANTS} is well suited to evaluate both the ideal equilibrium and the equilibrium produced by the filamentary coils. In both cases, 5000 particles are included in the evaluation for each flux surface for 200 ms. Previous calculations indicated that 5000 particles per flux surface were sufficient for Monte-Carlo statistical purposes \citep{bader2019stellarator}. These calculations only consider the vacuum fields, and improvements with finite pressure and plasma currents are left for future optimization studies. Similarly, calculations including collisional effects are left for future analyses.

\begin{figure}
  \centering
  \includegraphics[width=\textwidth]{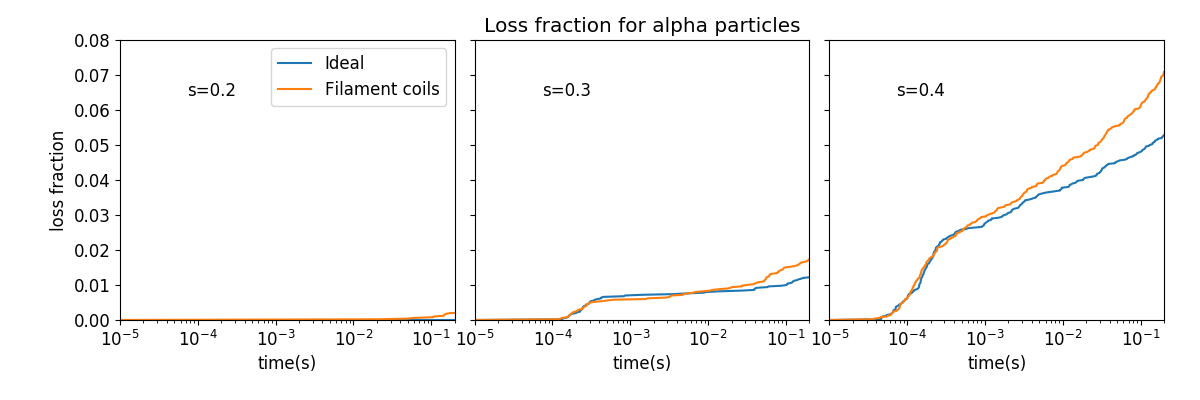}
  \caption{Collisionless alpha particle losses for an ARIES-CS scale device. Alpha particle losses are plotted as a function of time for three flux surfaces corresponding to normalized toroidal fluxes of 0.2 (left), 0.3 (center) and 0.4 (right). Blue represents the ideal optimized fixed boundary configuration. Orange represents the vacuum field produced by filamentary coils.}
\label{fig:losscoils}
\end{figure}

The results for the energetic particle confinement are shown both with and without coils in figure \ref{fig:losscoils}.  The confinement does deteriorate slightly with the presence of coils. Some particles are lost at $s=0.2$ whereas in the ideal case, no particles are lost. The losses just outside the mid-radius, at $s=0.3$ increase from 1.2\% in the ideal case to 1.7\% in the configuration with coils.

\begin{figure}
  \centering
  \includegraphics[width=0.6\textwidth]{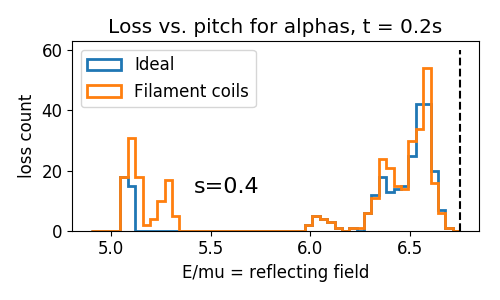}
  \caption{Alpha particle losses as a function of pitch angle for the ideal fixed boundary configuration (blue) and the vacuum field from filamentary coils (orange).  The dashed black line represents the trapped passing boundary.}
\label{fig:lvpcoils}
\end{figure}

The lost particles can be shown as a function of starting pitch angle at $s=0.4$ in figure \ref{fig:lvpcoils}. Here, the x-axis represents the field at which the particle reflects, $B_{ref} = E/\mu$ where $E$ and $\mu$ are the particle's energy and magnetic moment. Low values of $B_{ref}$ represent deeply trapped particles, and high values of $B_{ref}$ represent particles near the trapped-passing boundary. The trapped passing boundary is indicated with a vertical dashed line. All passing particles are confined. 

As is clear from the results in figure \ref{fig:lvpcoils}, most of the particle losses are from particles near the trapped-passing boundary. However, there are a few additional losses of deeply trapped particles,  and from particles somewhat further from the trapped passing boundary, at $E/\mu \approx 6.0$.

\subsection{Turbulent Transport}

Improving turbulent transport is a key area for stellarator research.  Recent results from W7-X indicate turbulent transport determines the overall energy and particle confinement \citep{pablant2020experiment}.  The dominance of turbulent transport in optimized stellarators had also been previously shown for HSX plasmas \citep{canik2007experiment}.  The configuration presented here is not explicitly optimized for turbulent transport. However, previous gyrokinetic calculations indicate that quasihelically symmetric configurations demonstrate enhanced nonlinear energy transfer properties over other optimized configurations \citep{plunk2017saturation,mckinney2019comparison}. It is anticipated that future iterations will include optimization of nonlinear turbulent energy transfer using a novel metric modeling turbulent energy transfer to stable modes \citep{hegna2018theory,faber2020ptsm3d}. Despite the lack of the turbulence metric in the optimization scheme, some aspects of the turbulence properties can be deduced from analyzing the equilibrium. 

\begin{figure}
  \centering
  \includegraphics[width=0.6\textwidth]{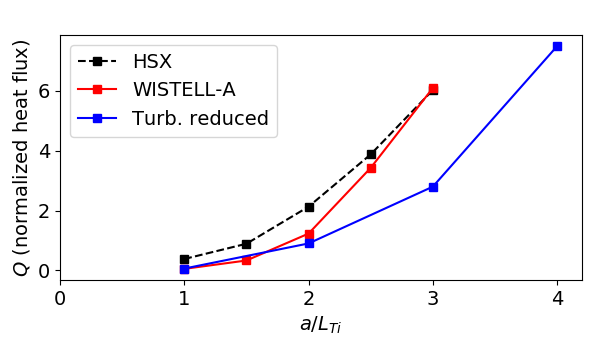}
  \caption{Turbulent heat flux at the $s=0.5$ surface as a function of normalized ion temperature scale length, $a/L_{Ti}$ for three different configurations, HSX (black dashed), WISTELL-A (red) and a turbulence reduced configuration (blue).}
\label{fig:turb}
\end{figure}

Non-linear flux-tube gyrokinetic calculations were performed to describe ion temperature gradient (ITG) turbulence at the $s=0.5$ surface using the \textsc{Gene} code \citep{jenko2000gene} for various values of the ion temperature scale length, $a/L_{Ti}$ assuming adiabatic electrons. In figure \ref{fig:turb}, the heat flux for WISTELL-A are compared to two other configurations, the HSX configuration which has been well-analyzed for turbulent transport \citep{faber2015pop,pueschel2016prl,faber2018jpp,mckinney2019comparison}, and a third ``turbulent reduced'' configuration, which was the result of a separate optimization calculation. The turbulence reduced configuration is identical to the configuration in \citep{bader2019stellarator} labeled "Opt. for QHS and $\Gamma_c$." It possessed favorable energetic particle properties, but did not possess a vacuum magnetic well, and thus was not considered as a viable configuration. Nevertheless, the turbulent properties of this configuration are of interest.  The results show that WISTELL-A reduces turbulent heat flux relative to HSX in the low ion temperature scale length regime. However, for $a/L_{Ti} \geq 2$,  the heat flux is comparable to that of HSX. The turbulence-reduced configuration, on the other hand, demonstrates reduced heat flux across a range of $a/L_{Ti}$ and a dramatic reduction in heat flux from HSX and WISTELL-A at $a/L_{Ti} \geq 3$.

\begin{figure}
  \centering
  \includegraphics[width=0.6\textwidth]{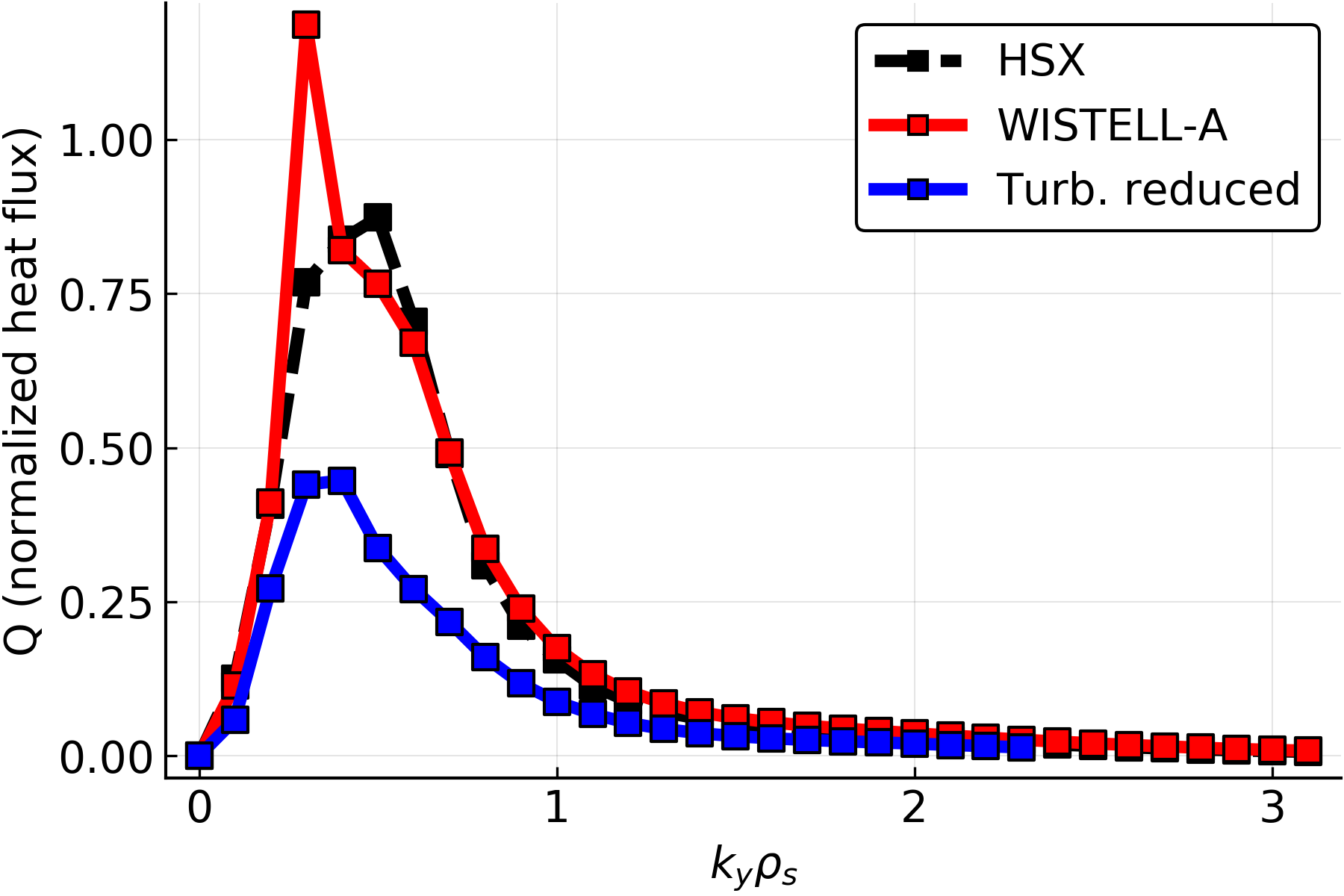}
  \caption{Turbulent heat flux spectrum at $a/L_{Ti}=3$ from Figure \ref{fig:turb} as function of normalized binormal wavenumber, $k_y\rho_s$ for HSX (black dashed), WISTELL-A (red) and a turbulence reduced configuration (blue).}
\label{fig:turb_flux_spectrum}
\end{figure}

\begin{figure}
    \centering
    \includegraphics[width=0.32\textwidth]{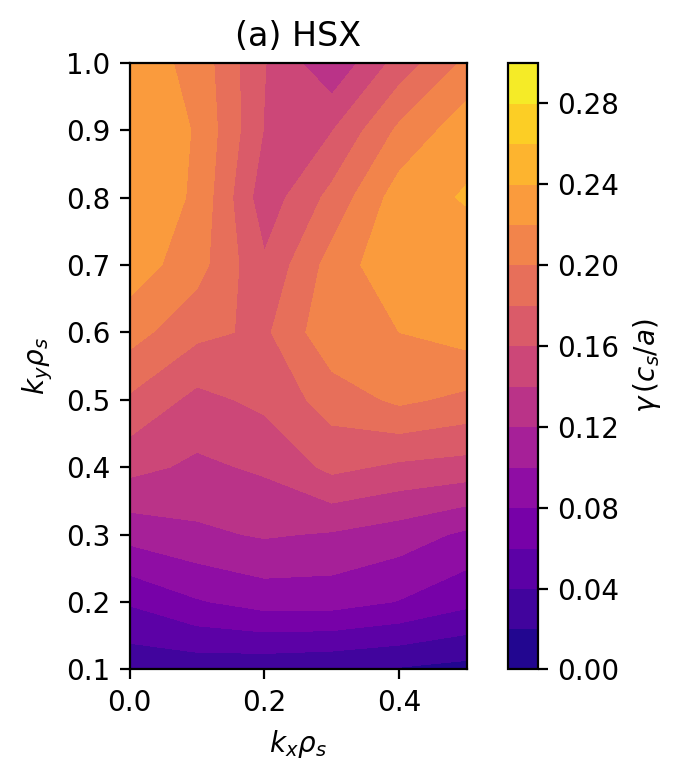}
    \includegraphics[width=0.32\textwidth]{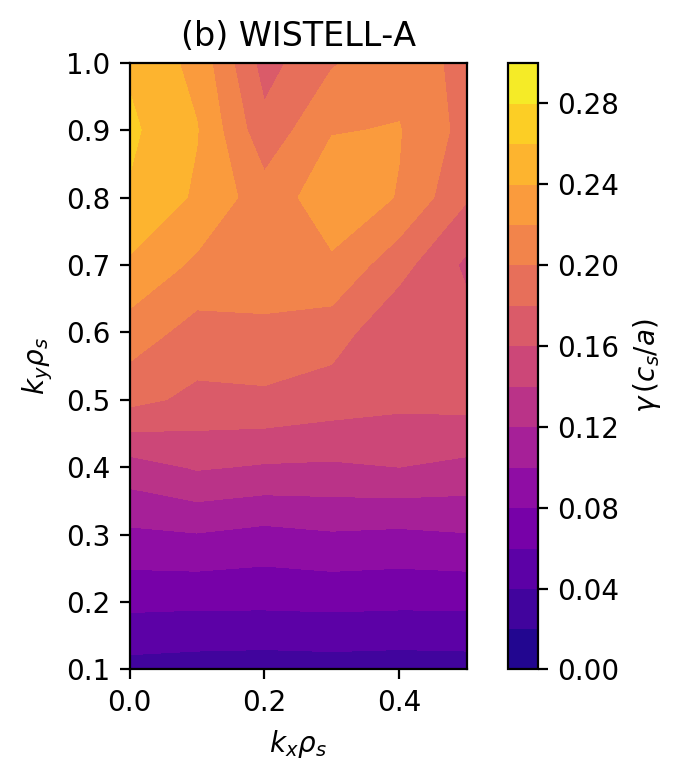}
    \includegraphics[width=0.32\textwidth]{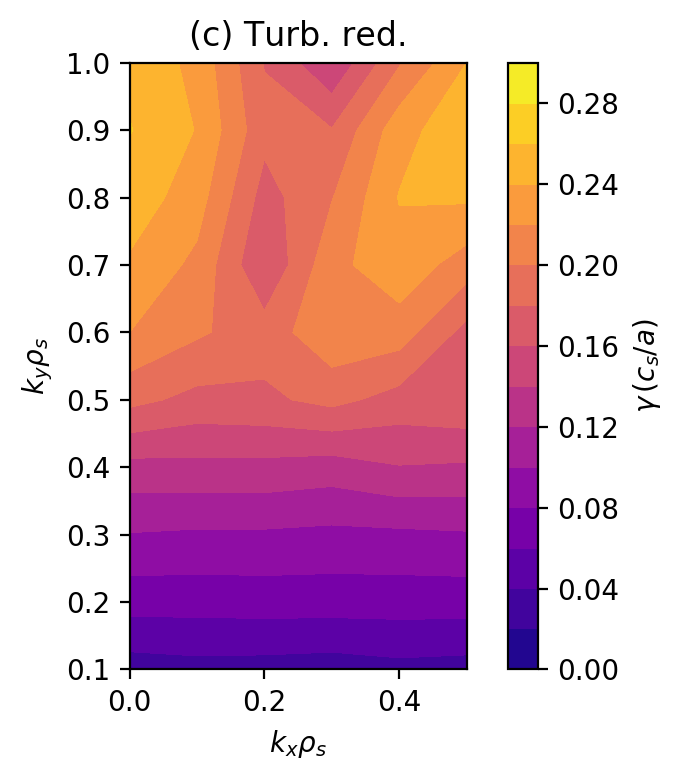}
    \caption{ITG growth rate spectrum with $a/L_{Ti} = 3$ as a function of normalized radial wavenumber $k_x\rho_s$ and normalized binormal wavenumber $k_y\rho_s$ (a) HSX, (b) WISTELL-A, and (c) turbulence-reduced configurations.}
    \label{fig:itg_growth_rates}
\end{figure}

Analysis of the turbulence at $a/L_{Ti} = 3$ for each configuration indicates that the differences in the ion heat flux values at $s = 0.5$ are not associated with changes in the linear ITG instability spectrum.  Figure \ref{fig:turb_flux_spectrum} shows the heat flux at $a/L_{Ti} = 3$ for each configuration as a function of $k_y\rho_s$.  The HSX and WISTELL-A show similar heat flux spectra, with the flux peaking at $k_y\rho_s \approx 0.6$.  The prominent feature in the WISTELL-A flux spectrum at $k_y\rho_s = 0.2$ has been previously observed and analyzed in gyrokinetic simulations of HSX \citep{faber2015pop,faber2018jpp} and while large in value, does not contribute substantially to the bulk of the heat flux.  More strikingly, the bulk of the heat flux spectrum in the turbulent reduced configuration is down-shifted in $k_y\rho_s$ from $k_y\rho_s \approx 0.6$ to $k_y\rho_s \approx 0.3$.  The linear growth rate spectra as a function of $(k_x\rho_s,k_y\rho_s)$ for each configuration at $a/L_{Ti} = 3$ is shown in figure \ref{fig:itg_growth_rates}.  Visual inspection of the growth rate spectra indicates there is little difference in the dominant linear instability between each configuration, and in fact, the turbulence-reduced configuration (figure \ref{fig:itg_growth_rates}c) has larger growth rates than either HSX or WISTELL-A.  This observation is supported more directly by using the eigenmode data from Fig.~\ref{fig:itg_growth_rates} in a quasilinear heat flux calculation using the model described in \citet[Eq. 2]{pueschel2016prl}, which is reproduced here:
\begin{equation}
    Q_{QL} = \frac{a}{L_{Ti}}\mathcal{C}\sum\limits_{k_x,k_y}\frac{w_i\left(k_x,k_y\right) \gamma\left(k_x,k_y\right)}{\left\langle k_{i,\perp}^2\left(k_x,k_y\right)\right\rangle},\,\,\,\left\langle k_\perp^2\left(k_x,k_y\right) \right\rangle = \frac{\int \mathrm{d}z \sqrt{g(z)} \Phi^2\left(k_x,k_y,z\right) k_\perp^2\left(k_x,k_y,z\right)}{\int \mathrm{d}z \sqrt{g(z)} \Phi^2(k_x,k_y,z)}.
\end{equation}
The linear growth rate $\gamma\left(k_x,k_y\right)$ is calculated from a linear \textsc{Gene} simulation at normalized perpendicular wavenumber $\left(k_x,k_y\right)$ and produces an eigenmode $\Phi\left(k_x,k_y,z\right)$, where $z$ is the field-line-following coordinate.  The Jacobian along the field line is given by $\sqrt{g(z)}$ and each contribution to the sum is weighted by $w_i = \tilde{Q}_i/\tilde{n}_i^2$, where $\tilde{Q}_i$ and $\tilde{n}_i$ are the calculated linear gyrokinetic heat flux and density perturbations from \textsc{Gene}.  The normalizing coefficient $\mathcal{C}$ is fit to a nonlinear gyrokinetic heat flux calculation and $a/L_{Ti}$ is the normalized temperature gradient; only ratios of $Q_{QL}$ will be considered here to avoid model ambiguity. The quasilinear calculation predicts the heat flux for the turbulence reduced configuration should actually be larger than for HSX at $a/L_{Ti}=3$ by a factor of approximately 1.1. This does not agree with the nonlinear gyrokinetic heat fluxes shown in Fig.~\ref{fig:turb}.  The discrepancy between the linear growth rates and the full nonlinear heat flux is consistent with trends found in \citet{mckinney2019comparison}. The results presented here also indicate that linear growth rates can be a misleading indicator for stellarator turbulence and turbulent transport. Furthermore, these results suggest that the turbulence-reduced configuration possesses enhanced turbulence saturation mechanisms.

\begin{figure}
    \centering
    \includegraphics[width=0.48\textwidth]{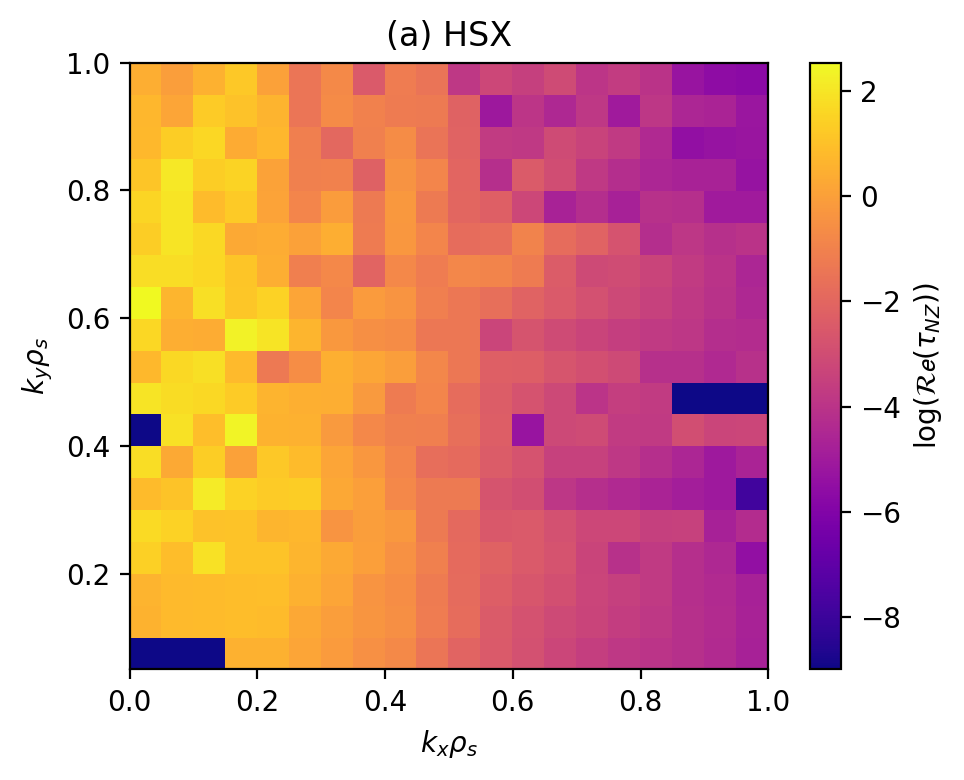}
    \includegraphics[width=0.48\textwidth]{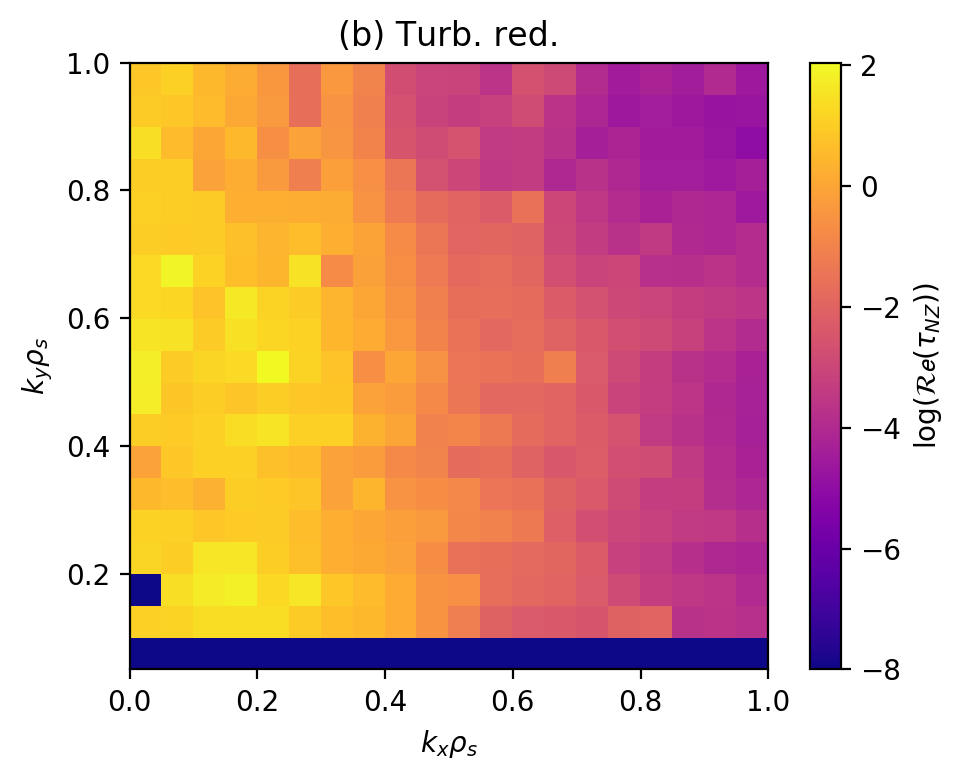}
    \caption{Triplet correlation lifetimes on a log scale as a function of $\mathbf{k}$ for (a) HSX and (b) the turbulence-reduced configuration for $a/L_{Ti}=3$.  The value shown at any particular $\mathbf{k}$ is the calculation of $\tau_{pst}$ as defined by \ref{eq:triplet_tau} where $p$ is an unstable mode at that $\mathbf{k}$, $s$ is a stable mode at \emph{whatever} wavenumber $\mathbf{k}'$ such that $\mathrm{Re}\left(\tau_{pst}\left(\mathbf{k},\mathbf{k}'\right)\right)$ is maximized by a third mode $t$, which can be unstable or stable.  The triplet correlation lifetime value has been weighted by a fluctuation energy spectrum obtained from gyrokinetic calculations which emphasize triplet lifetimes involving energy containing scales. }
    \label{fig:triplet_lifetimes}
\end{figure}

To make a preliminary assessment of the turbulence saturation characteristics, the turbulence saturation theory from \citet{hegna2018theory} will be applied. A crucial aspect of the this theory is the supposition that the dominant nonlinear physics involves energy transfer from unstable to damped eigenmodes at comparable wavenumbers.  This is accomplished through a three-wave interaction quantified by a triplet correlation lifetime between unstable and stable ITG modes as defined in \citet[Eq. 104]{hegna2018theory} by
\begin{equation}
    \tau_{pst}\left(\mathbf{k},\mathbf{k'}\right) = \frac{-\mathrm{i}}{\omega_t\left(\mathbf{k}''\right) + \omega_s\left(\mathbf{k}'\right) - \omega^\ast_p\left(\mathbf{k}\right)};\,\,\,\,\mathbf{k} -\mathbf{k}' = \mathbf{k}'',
    \label{eq:triplet_tau}
\end{equation}
where $\omega(\mathbf{k})$ is the complex linear ITG frequency at normalized wavenumber $\mathbf{k} = \left(k_x\rho_s,k_y\rho_s\right)$.  Large values of the triplet lifetimes suggest energy can be very effectively transferred out of turbulent-transport-inducing instabilities into damped eigenmodes that either dissipate energy or transfer it back to the bulk distribution function.  High values of $\tau_{pst}$ correspond to lowered turbulent fluctuation levels and correspondingly reduced turbulent transport.   In figure \ref{fig:triplet_lifetimes}, the triplet correlation lifetimes are shown for the HSX and the turbulence optimized configuration.  Importantly, the turbulence-reduced configuration shows larger triplet correlation lifetimes in the region $k_y\rho_s \lesssim 0.6$ compared to HSX, where the larger correlation lifetimes are observed at higher $k_y\rho_s$.  This is an important difference, as instabilities at larger scales (smaller $|\mathbf{k}_\perp|$) can more easily contribute to turbulent transport.  Thus, larger triplet correlation lifetimes at smaller $k_y\rho_s$ where at least one unstable and one stable mode are involved suggests energy is being transferred more efficiently from the modes driving the fluctuation spectrum to dissipation and thus lowering the contribution to turbulent transport at that $k_y\rho_s$. This may be contributing to both the decrease in overall transport in figure \ref{fig:turb} and the downshift in heat flux spectrum in figure \ref{fig:turb_flux_spectrum} between HSX and the turbulence-reduced configuration.  The connections between nonlinear turbulent heat flux, quasilinear turbulent heat flux, and the triplet correlation lifetimes are summarized in table \ref{tab:turb}.  The triplet correlation lifetimes for each configuration are quantified by computing a spectral average defined as
\begin{equation}
    \left\langle \tau_{NZ} \right\rangle = \sum\limits_{k_x,k_y} \mathcal{S}_{G}\left(k_x,k_y\right) \mathrm{Re}\left(\tau_{NZ}\left(k_x,k_y\right)\right).
\end{equation}
The weighting factor $\mathcal{S}\left(k_x,k_y\right)$ is a turbulent fluctuation spectrum computed from a characteristic nonlinear gyrokinetic simulation that preferentially weights low $|\mathbf{k}_\perp|$ contributions to provide consistency with the nonlinear simulations. The decrease in nonlinear flux between HSX and the turbulence reduced configuration by a factor of 2 correlates with an increase in triplet correlation lifetimes by more than a factor of two, while the increase in nonlinear flux between WISTELL-A and HSX correlates with a decrease in triplet correlation lifetimes. This will be explored in more detail in future work, but this result already indicates that quasihelically symmetric configurations with lower ITG-driven transport can be obtained.

\begin{table}
    \centering
    \begin{tabular}{c|c|c|c}
         Config. & $Q_{NL}/Q_{NL,HSX}$ & $Q_{QL}/Q_{QL,HSX}$ & $\left\langle\tau_{NZ}\right\rangle/\left\langle\tau_{NZ,HSX}\right\rangle$ \\ \hline
         HSX & 1 & 1 & 1 \\
         WISTELL-A & 1.1 & 1.05 & 0.76 \\         
         Turb. red. & 0.5 & 1.1 & 2.85
    \end{tabular}
    \caption{Values of nonlinear gyrokinetic heat flux, quasi-linear heat flux, and spectral averaged triplet correlation lifetimes at $a/L_{Ti} = 3$ for the three configurations.  All values have been normalized to the corresponding HSX value.}
    \label{tab:turb}
\end{table}

\subsection{MHD Stability}
\label{sec:mhd}

The MHD properties of QH stellarators \citep{nuhrenberg1988qhs} are somewhat distinct from other classes of optimized stellarators.  The relatively reduced connection length (the distance along the field line between $B_{max}$ and $B_{min}$) implies that QH stellarators have reduced banana widths, reduced orbit drifts for passing particles \citep{talmadge2001orbits}, smaller Pfrisch-Schl\"uter \citep{boozer1981coordinates, schmitt2013ps} and bootstrap currents \citep{boozer1990BOOTSJ, schmitt2014bs} and reduced Shafranov shift than an equivalent sized tokamak at the same parameters.  This is quantified by the ``effective" rotational transform $\stkout{\iota}_{eff} = (\stkout{\iota} - N)$ where $N$ is the periodicity  of the stellarator.  Moreover, the bootstrap current in a QH stellarator is in the opposite direction  relative to what occurs in a tokamak.  This has the consequence of reducing the value of $\stkout{\iota}$ with rising plasma pressure and producing negative $d\stkout{\iota}/ds$ in the core region.  Negative values of $d\stkout{\iota}/ds$ can have beneficial effects for both ideal ballooning \citep{hegna2001ballooning} and magnetic island physics \citep{hegna1994islands}. 

The only stability quantity constrained in the ROSE optimization  is the magnetic well depth as described by $d^2V/d\Phi^2$, the second derivative of volume with respect to toroidal flux, at the magnetic axis for the vacuum equilibrium. A magnetic well ($V'' < 0$) is necessary for stability against interchange modes. The magnetic well depth at points away from the magnetic axis was not explicitly optimized for, but can be quantified with

\begin{equation}
   W =  \left(\left.\frac{dV}{d\Phi}\right|_{\rho=0} - \frac{dV}{d\Phi}\right)/\left. \frac{dV}{d\Phi}\right|_{\rho=0}
    \label{eq:well}
\end{equation}

\begin{figure}
  \centering
  \includegraphics[width=1.0\textwidth]{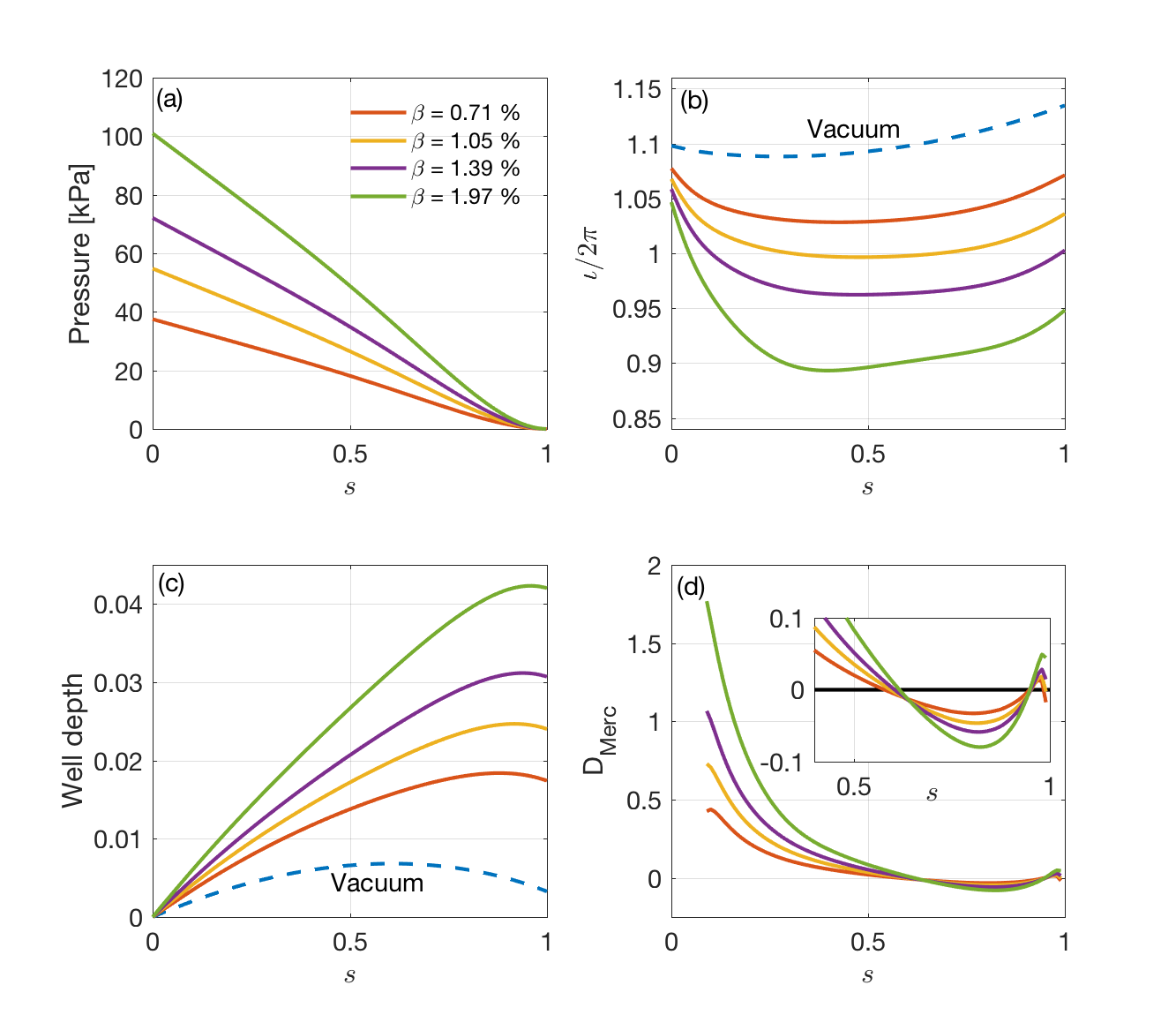}
  \caption{(a) The pressure profiles versus normalized flux, $s$, for several values of volume-average $\beta$. 
  The effects of finite-$\beta$ are shown for the radial profiles (in $s$) of the (b) rotational transform ($\stkout{\iota} \equiv \iota / 2\pi$), (c) the magnetic well depth and (d) the Mercier stability criterion. Positive values indicate Mercier stability. (inset shows more detail for $0.4<s<1.0$.) Vacuum quantites for the rotational transform and well depth are shown (dashed lines).}
\label{fig:beta_scan_part_1}
\end{figure}

For calculations including finite pressure, a pressure profile is assumed where temperature is linear in normalized flux, $T = T_0\left(1-s\right)$ and the density profile is broad, $n = n_0\left(1 - s^5\right)$.  Here, $T_0$ and $n_0$ represent the temperature and density at the magnetic axis respectively. The pressure profiles as a function of normalized flux are given in figure \ref{fig:beta_scan_part_1}a. The pressure was varied by varying $T_0$ at fixed $n_0$ = 0.9 $\times 10^{20} m^{-3}$, with $T_0$ ranging from 1.3 keV to 3.5 keV.  The free-boundary equilibrium was calculated with VMEC using the vacuum magnetic field given by the filamentary coils described in section \ref{sec:coils}.  The self-consistent bootstrap current profiles are calculated using SFINCS \citep{landreman2014sfincs} in an iterative loop with the VMEC equilibrium. In the neoclassical calculations, a pure plasma with $T_e = T_i$ was assumed, and the bootstrap current was calculated at the ambipolar radial electric field.  The rotational transform profiles from the VMEC equilibria evaluated for several different pressures are shown in figure \ref{fig:beta_scan_part_1}b.

As noted previously, the bootstrap current tends to lower the value of $\stkout{\iota}$ and produce reversed magnetic shear in the core.  As seen in figure \ref{fig:beta_scan_part_1}b, the rotational transform profile crosses $\stkout{\iota}=1$ around $s \approx 0.5$ when the normalized pressure, $\beta \approx 1\%$. Unless compensated for, this potentially sets an operational limit for this configuration. 

With finite beta equilibria, relevant stability metrics can be calculated. The well depths, as given by equation \ref{eq:well} are shown in figure \ref{fig:beta_scan_part_1}c. As seen in the figure, the vacuum configuration has a magnetic well, and the well depth gets larger as the pressure increases.  However, a magnetic hill region remains near the plasma edge.  

The Mercier criterion is given by the sum \citep{bauer1984mhd, carreras1998torsatronconfigs}:
\begin{equation}
    \label{eq:mercier}
    D_{Merc} = D_S + D_W + D_I + D_G \geq 0
\end{equation}
where the individual terms in eq. \ref{eq:mercier} represent contributions (stabilizing or destabilizing) from the shear, magnetic well, current and geodesic curvature and are given by the following:
\begin{align*}
   D_S  &= \frac{s}{\stkout{\iota}^2 \pi^2} \frac{\left(\Psi'' \Phi'\right)^2}{4} \\
   D_W  &= \frac{s}{\stkout{\iota}^2 \pi^2} \int \int {g d\theta d\zeta  \frac{B^2}{g^{ss}} \frac{dp}{ds}} \times \left(V'' - \frac{dp}{ds}\int \int{g \frac{d\theta d\zeta}{B^2}} \right) \\
   D_I  &= \frac{s}{\stkout{\iota}^2 \pi^2}\left[\int \int{g d\theta d\zeta  \frac{B^2}{g^{ss}}\Psi'' I'} - \left(\Psi'' \Phi' \right) \int \int{g d\theta d\zeta \frac{\left(\Vec{J}\cdot\Vec{B}\right)}{g^{ss}}}  \right] \\
   D_G  &= \frac{s}{\stkout{\iota}^2 \pi^2}\left[ \left( \int \int{g d\theta d\zeta \frac{\left(\Vec{J}\cdot\Vec{B}\right)}{g^{ss}}} \right)^2  - \left(\int \int{g d\theta d\zeta \frac{\left(\Vec{J}\cdot\Vec{B}\right)^2}{g^{ss} B^2} }\right)\left(\int \int { g d\theta d\zeta  \frac{B^2}{g^{ss}}} \right) \right]\\
\end{align*}

In the above expressions, $\Phi$ and $\Psi$ are the toroidal and poloidal magnetic fluxes, $g$ is the Jacobian, $p$ is the pressure, $I$ is the net toroidal current enclosed within a magnetic surface, and the metric element $g^{ss} = |\nabla s|^2$.

\begin{figure}
  \centering
  \includegraphics[width=1.0\textwidth]{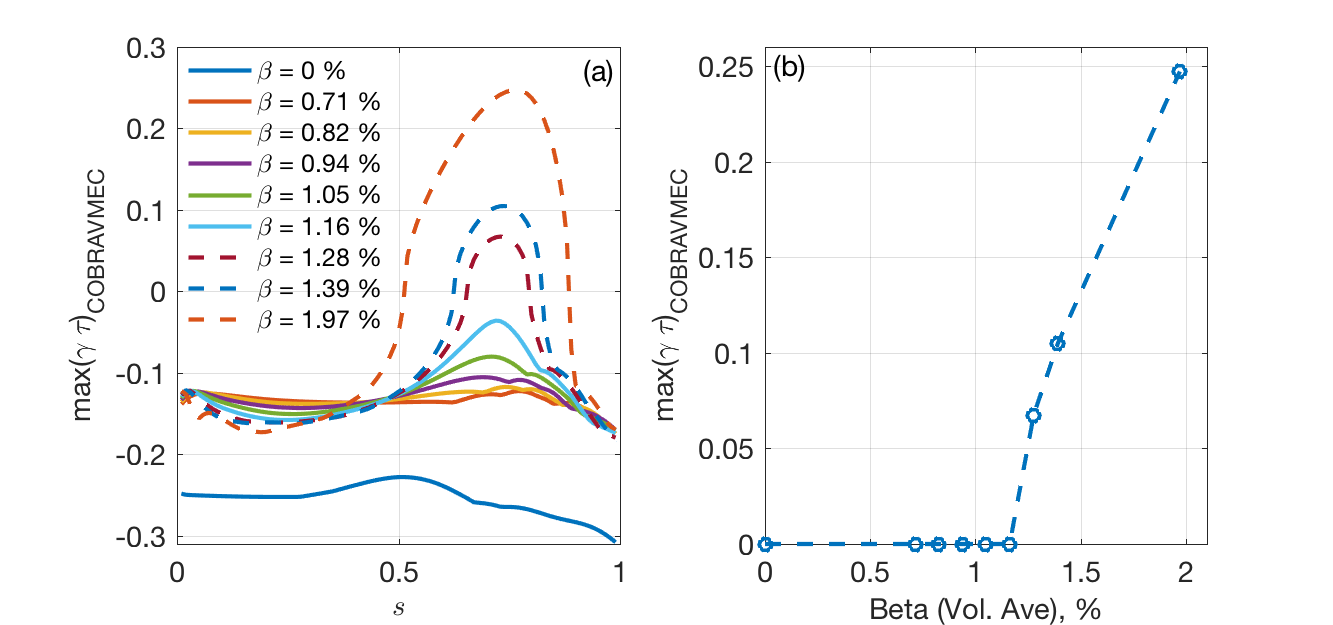}
  \caption{(a) The radial profiles of the of the growth rates, as calculated by COBRAVMEC for various values of normalized pressure, $\beta$. (b) The value of the growth rate for the most unstable ballooning mode is shown as a function of normalized pressure, $\beta$.  The configuration at $\beta$=1.16\% is stable to ballooning modes.}
\label{fig:cobravmec}
\end{figure}

Figure \ref{fig:beta_scan_part_1}d shows the Mercier stability criterion as given by eq. \ref{eq:mercier} and evaluated by VMEC. All configurations are Mercier stable ($D_{Merc} > 0$) for $s \lesssim 0.6$. Calculations of ballooning stability were obtained with COBRAVMEC \citep{sanchez2000cobra} and are shown in figure \ref{fig:cobravmec}. The configuration is stable to ballooning modes up to values of $\beta \leq 1.2\%$.  Ballooning stability is violated at higher values of $\beta$ with the specified pressure profile shape.  The region where ballooning instability tends to occur first is near $s \approx 0.7$. Higher critical $\beta$ values for ideal ballooning instability can be obtained by tuning the pressure profile.  This will be pursued in future work.

While the configuration presented here has only modest MHD stability properties, it is anticipated these properties can be improved through further optimization.  However, there is little evidence to support the notion that MHD stability provides any rigorous limit for stellarator operation \citep{weller2001MHDstability}.  Rather, MHD equilibrium properties are thought to provide a more stringent limit on plasma $\beta$.

\subsection{Divertor}
\label{sec:div}

Divertors for quasisymmetric stellarators require either resilience to changes in the plasma current and pressure profiles, or active control mechanisms to ensure proper function of a resonant divertor, often referred to as an island divertor, from startup to the operational point \citep{konig2002divertor}. Properties for non-resonant divertors have been explored both theoretically and numerically \citep{boozer2018simulation, bader2017hsx, bader2018minimum}. In this section we present a methodology for constructing such a divertor, and provide a first attempt at what a non-resonant divertor could look like for a quasihelically symmetric stellarator.

The methodology for constructing the divertor is to begin with a wall at some uniform distance from the last closed flux surface. A field line diffusion model can then be used to calculate the strike positions on the wall \citep{strumberger1992magnetic, bader2017hsx}. The model works by distributing field lines uniformly on good internal flux surfaces. Points are followed along the field line but given a random perpendicular displacement in accordance with a specified diffusion parameter. The "diffusive" field lines eventually leave the confined region and terminate on the wall. These exiting field lines are always seen to exit the plasma in regions of high curvature of the last closed flux surface \citep{strumberger1992magnetic, bader2017hsx}. 

As noted above, a benefit of the \texttt{FOCUS} code is that it allows the coils to expand away from the plasma in regions where they are not required to be close to the plasma. Fortuitously, the regions of high curvature where strike lines exit are also regions where coil expansion is possible. Therefore, a method for divertor construction is to adjust the uniform wall so it is expanded in these regions. This allows for both longer connection lengths between the plasma and the wall, and some degree of divertor closure, allowing for access to high neutral pressure. 

\begin{figure}
  \centering
  \includegraphics[width=\textwidth]{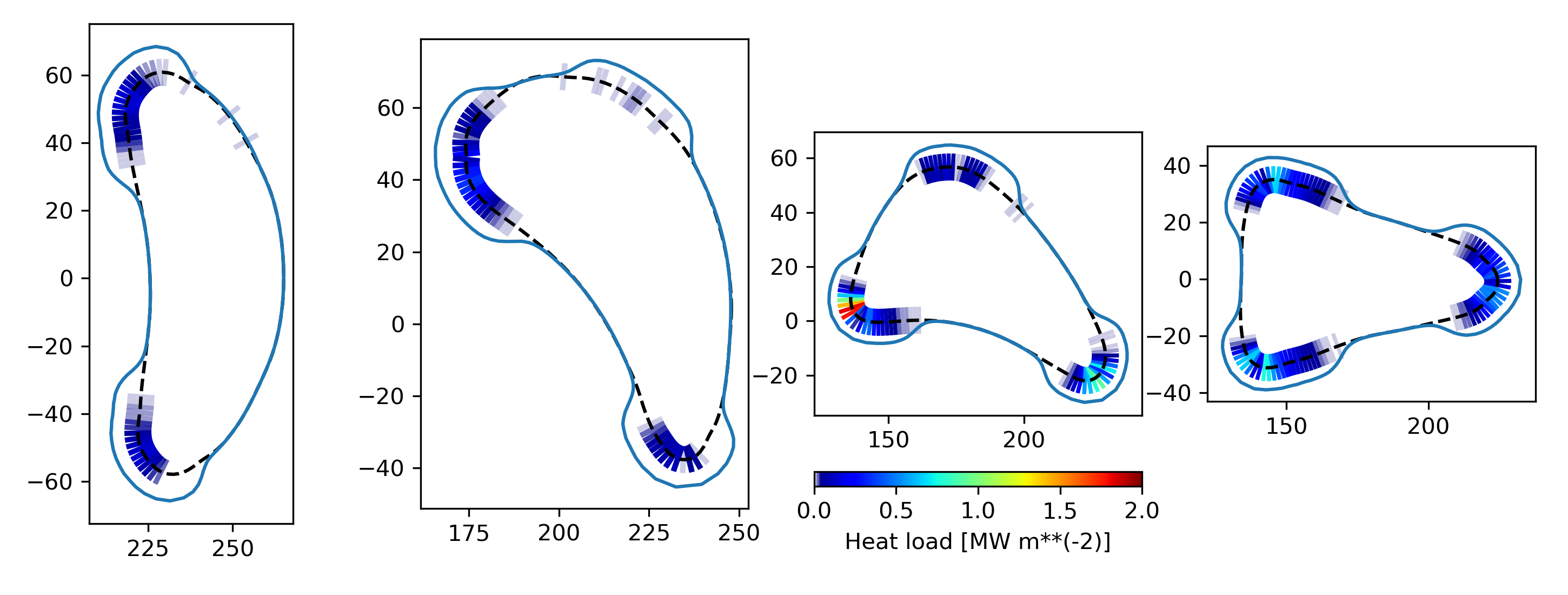}
  \caption{A representation of a non-resonant divertor concept for the WISTELL-A device. Plots correspond to four toroidal positions at $\phi$ = 0$^\circ$, 15$^\circ$, 30$^\circ$ and 45$^\circ$. The wall (solid blue) is expanded near regions of peak heat flux. The heat flux is calculated by EMC3-EIRENE and shown in color.}
\label{fig:divcons}
\end{figure}

\begin{figure}
  \centering
  \includegraphics[width=0.8\textwidth]{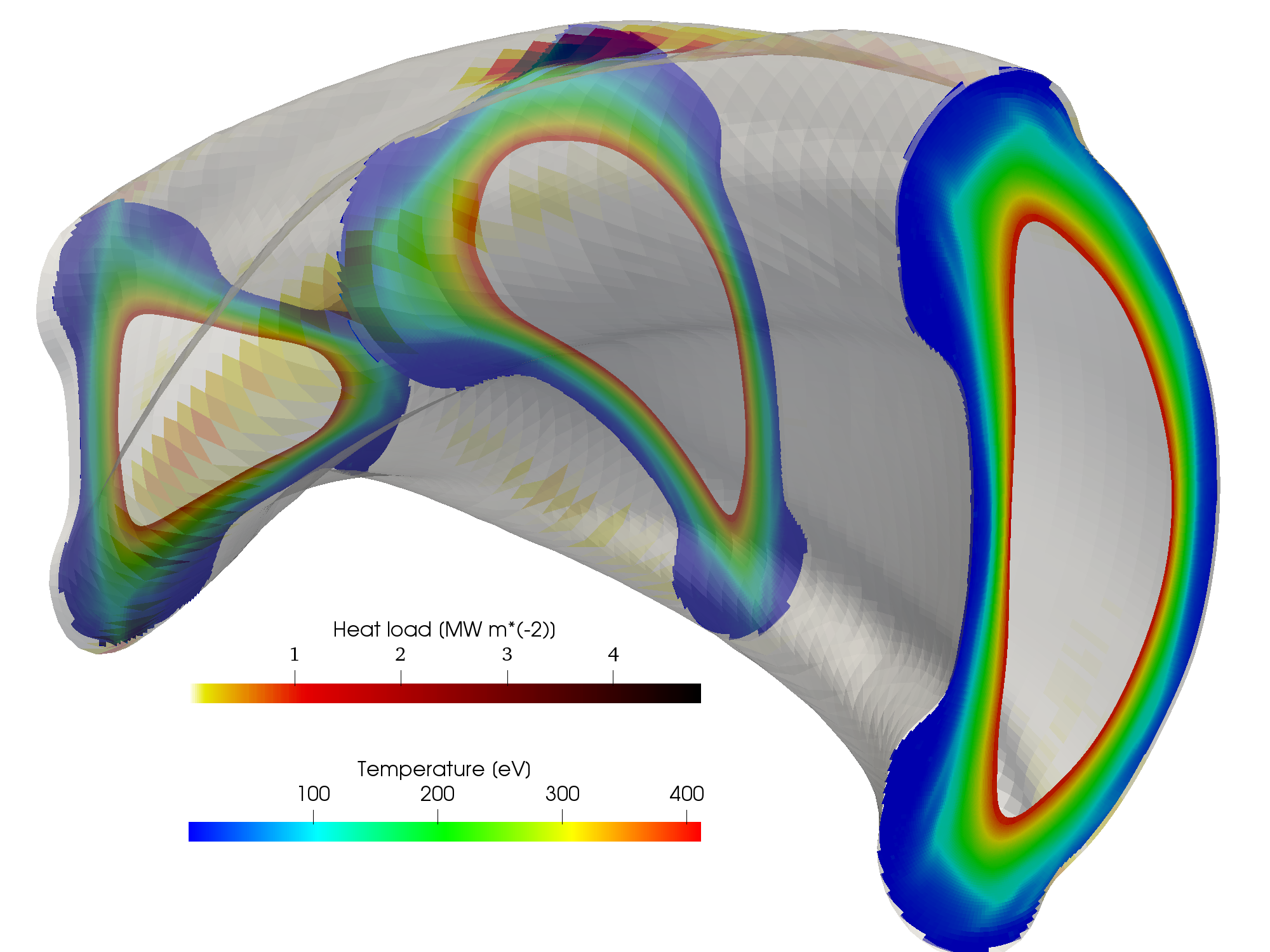}
  \caption{A three dimensional representation of a non-resonant divertor concept for the WISTELL-A device. The gray outer surface represents the wall. Temperature contours from EMC3-EIRENE are presented at $\phi = $ 0$^\circ$, 22.5$^\circ$ and 45$^\circ$, and the heat flux on the wall is also represented on the boundary.}
\label{fig:div3d}
\end{figure}

An initial attempt at such a construction is shown in figure \ref{fig:divcons}. Here four plots of the divertor structure at toroidal values of $\phi$ = 0$^\circ$, 15$^\circ$, 30$^\circ$ and 45$^\circ$ are shown. In addition, an EMC3-EIRENE simulation was carried out with nominal operating parameters for the upgraded scenario. The calculation from EMC3-EIRENE indicates that the heat flux is concentrated in specific areas toroidally and poloidally near $\phi$ = 30$^\circ$. A three dimensional representation of the divertor design is shown in figure \ref{fig:div3d}

Future iterations of the divertor structure design are necessary in order to smooth the heat flux deposition. The results presented here are therefore meant to indicate a first attempt at how divertor design could proceed and not indicative of a final design.

In addition, the configuration will have access to island divertor experiments by exploiting the $n=8,m=7$ resonance. However, due to the presence of self-generated plasma currents in quasisymmetric equilibria, ensuring the island position is maintained throughout the discharge to the operating point requires some external control, whether by auxiliary coils or current drive. Designing such operational scenarios are beyond the scope of this current work. 

\section{Conclusion}
\label{sec:conc}

A new optimized quasihelically symmetric stellarator is developed that has a number of desirable features including improved energetic ion confinement, low neoclassical transport, reduced turbulent transport and non-resonant divertor capability.  This configuration is made possible through the simultaneous improvements in optimization (ROSE) and coil (FOCUS) tools as well as advances in physics understanding.   Stellarator optimization is a rapidly developing field with new advancements in physics metrics, equilibrium solutions and optimization algorithms occurring at an impressive pace. Individual optimized equilibria represent markers of where progress is at a given point in time. They highlight recent advancements and provide benchmarks for testing further optimizations. 

A particularly significant advance is the development of configurations with excellent energetic particle confinement.  Importantly, this improved energetic ion confinement is realized in the presence of external magnetic fields generated from filamentary coils.  These improvements were only possible due to recent improvements in stellarator optimization.

A relatively new element in stellarator optimization is targeting reduced turbulent transport. In addition to the configuration presented here, another turbulence-reduced configuration has been identified.  Nonlinear gyrokinetic simulations of ion temperature gradient turbulence demonstrate that this configuration has improved normalized values of heat flux relative to that predicted in HSX for all values of ion temperature gradient drive.  Interestingly, improvements in turbulent transport are not correlated to significant changes to the linear stability properties.  Rather, there are indications that the improved turbulent transport is related to changes in the nonlinear energy transfer physics.

Quasihelically symmetric stellarators have a number of intrinsic advantages when compared to other optimized stellarators.  The 
reduced connection lengths of QH relative to QA implies they have smaller banana widths, Shafranov shifts, and plasma generated currents. However, MHD effects need to be accounted for in comprehensive stellarator optimization.  Only modest attempts were made to improve the MHD properties of this new configuration. Nevertheless, configurations with magnetic wells throughout the core confinement regime are realized with ballooning instability onset at a few percent $\beta$.   There is reason to believe this property can be improved.  

There is a need for a viable divertor option for stellarators with finite bootstrap current. Initial calculations show that non-resonant divertors, which do not rely on a low-order resonance at the edge, may be a possible solution. The complex interaction between edge plasma, impurities, neutral gas, and plasma boundary surfaces in a stochastic edge are only accessible numerically with the EMC3-EIRENE code. Validation of this code in different edge scenarios in experiments with relevant geometries and conditions is necessary to predict the functionality of next step experiments.

A final advantage of new equilibria is that they can be the foundation for new experimental designs. In the appendix of this paper, a conception of a midscale quasihelically-symmetric stellarator is given.

\section{Acknowledgments}

This work was supported by the University of Wisconsin, UW2020-135AAD3116 and the US Department of energy grants DE-FG02-93ER54222 and DE-FG02-99ER54546. BJF is supported by the U.S. Department of Energy Fusion Energy Sciences Postdoctoral Research Program administered by the Oak Ridge Institute for Science and Education (ORISE) for the DOE. ORISE is managed by Oak Ridge Associated Universities (ORAU) under DOE contract number DE-SC0014664. All opinions expressed in this paper are the author's and do not necessarily reflect the policies and views of DOE, ORAU, or ORISE. This research used resources of the National Energy Research Scientific Computing Center (NERSC), a U.S. Department of Energy Office of Science User Facility operated under Contract No. DE-AC02-05CH11231.

\FloatBarrier

\appendix

\section{}\label{appA}

This appendix details a possible realization of the new configuration as a midscale experiment. Such an experiment could significantly advance the quasisymmetric concept and retire some of the risks related to energetic particle transport, turbulent transport and divertor operation in quasisymmetry. 

A 0-Dimensional analysis was carried out in order to determine a target equilibrium for a realization of the WISTELL-A configuration for a midscale experiment. Details are presented in table \ref{tab:0d}. The major size and cost drivers are the magnetic field strength and the minor radius. In addition the operation is split into two phases, an initial operational phase at half field (1.25 T), and a full operational phase at 2.5 T. The magnetic field strength is chosen to take advantage of Electron Cyclotron Heating (ECH) from commercially available 70 GHz gyrotrons at the second harmonic in the initial phase. In the full operational phase, the 70 GHz gyrotrons can be used at the fundamental o-mode harmonic, or 140 GHz gyrotrons can be used to heat at the second harmonic. The choice of gyrotron frequency sets the density cut-off, the maximum operational density for ECH plasmas. The cutoff density is $\epsilon_0 m_e \omega^2 / e^2$, where $\epsilon_0$ is the permittivity of free space, $m_e$ is the electron mass, $e$ is the fundamental charge and $\omega$ is the angular frequency of the launched wave. For 70 GHz gyrotrons the cutoff is $\sim$ 0.6 $\times$ 10$^{20}$ m$^{-3}$.  For 140 GHz gyrotrons the cutoff is $\sim$ 2.4 $\times$ 10$^{20}$ m$^{-3}$. The Sudo density limit for 1 MW absorbed power is 9.0 $\times$ 10$^{20}$ m$^{-3}$ \citep{sudo1990scalings}.

\begin{table}
\begin{center}
\def~{\hphantom{0}}
\begin{tabular}{lcccc}
     \hline
     \textbf{Param.} & \textbf{Initial} & \textbf{Upg.($H$=1)}  & \textbf{Upg.($H$=1.5)}  & \textbf{Upg.($H$=2)} \\
     \hline
     \hline
     $R$(m) & \multicolumn{3}{c}{2.0}\\
     $a$(m) & \multicolumn{3}{c}{0.3}\\
     $V$(m$^{3}$) & \multicolumn{3}{c}{3.55}\\
     $\stkout{\iota}$ & \multicolumn{3}{c}{1.1}\\
     $B$(T) & 1.25 & 2.5 &2.5 & 2.5\\
     ECH (MW) & 0.5 & 1.0 & 1.0 & 1.0\\
     NBI (MW) & 0.0 & 1.0 & 1.0 & 1.0\\
     $H$ factor & 1.5 & 1.0 & 1.5 & 2.0 \\
     $n$ (10$^{20}$ m$^{-3}$) & 0.15 & 0.9 & 0.9 & 0.9\\
     \hline
     $T_e$ (keV) & 1.2 & 0.8 & 1.3 & 1.7\\
     $T_i$ (keV) & 0.7 & 0.8 & 1.3 & 1.7  \\
     $\beta$ \% & 0.36 & 0.49 & 0.73 & 0.98 \\
     $\nu^*_i$ & 0.13 & 0.5 & 0.22 & 0.12 \\
     $\tau_E$ (ms) & 48 & 65 & 98 & 130 \\
     $\tau_{ie}$ (ms) & 35 & 5 & 9 & 14 \\
     \hline

\end{tabular}
\end{center}
\caption{Parameters and derived quantities (using 0-D analysis) for the WISTELL-A stellarator for the initial operational phase and 3 scenarios for the full operational phase}
\label{tab:0d}
\end{table}

The confinement time $\tau_E$ (in ms) is given by the ISS04 empirical scaling law \citep{yamada2005characterization},
\begin{equation}
    \tau_E = \tau_E^{\mathrm{ISS04}} = 134 a^{2.28}R^{0.64}P^{-0.61}n_e^{0.54}B^{0.84} \stkout{\iota}_{2/3}^{0.41}.
\end{equation}
Here, $R$ and $a$ are the major and minor radii respectively in meters, $P$ is the total absorbed power in MW, $n_e$ is the electron density in units of 10$^{19}$ m$^{-3}$, $B$ is the magnetic field on axis in T and $\stkout{\iota}_{2/3}$ is the rotational transform value at $r/a$ = 2/3. While the convention for the ISS04 scaling is to use density in units of 10$^{19}$ m$^{-3}$, from this point forward, all calculations will use the convention of density in units of 10$^{20}$ m$^{-3}$.  The average temperature, $T = \left( T_e + T_i \right)/2$ in eV is given by,
\begin{equation}
    T = \frac{\tau_e\left( P_i + P_e \right)}{3nV}
\end{equation}
where $V$ is the plasma volume, $P_i$ is the power absorbed by ions (taken here as power from the neutral beam) and $P_e$ is the power transmitted to the electrons. (taken here as ECH power).  In the table a confinement improvement factor $H$ is included anticipating potential advances in understanding how to reduce turbulent transport.  

The electron-ion energy equilibration time, $\tau_{ie}$ is
\begin{equation}
    \tau_{ie} = \frac{m_i}{2m_e}\tau_e;\;\tau_e = 3\left(2 \pi\right)^{3/2}\frac{\epsilon_0 m_e^{1/2} T^{3/2}}{n Z^2 e^4 \mathrm{ln}\Lambda}
\end{equation}

Here, $\epsilon_0$ is the permittivity of free space, $T$ is the mean temperature in joules, $n$ is the density, $Z$ is the particle charge (taken to be 1), $e$ is the fundamental electric charge, $m_e$ and $m_i$ are the electron and ion masses respectively, and $\mathrm{ln}\Lambda$ is the coulomb logarithm, taken to be 17. For the values calculated in table \ref{tab:0d}, we assume main species hydrogen with $n = n_e = n_i$. 

The energy partition between ions and electrons is calculated assuming,

\begin{equation}
    \frac{2}{3} \frac{P_e}{nV} = \left[ \frac{1}{\tau_E} + \frac{1}{\tau_{ie}}\right]T_e - \frac{T_i}{\tau_{ie}};\;
    \frac{2}{3} \frac{P_i}{nV} = \left[ \frac{1}{\tau_E} + \frac{1}{\tau_{ie}}\right]T_i - \frac{T_e}{\tau_{ie}}
\end{equation}
Where, ion and electron temperature, $T_e$ and $T_i$ respectively are in joules.

The normalized ion collisionality, $\nu^*_i$ is given by,
\begin{equation}
    \nu^*_i = \frac{1}{\tau_{ii}} \sqrt{\frac{m_i}{T_i}} \frac{R}{\epsilon^{3/2}\left(N-\stkout{\iota}\right)};\; \tau_{ii} = \frac{12\pi^{3/2}}{\sqrt{2}}\frac{m_i^{1/2} T_i^{3/2} \epsilon_0^2}{n Z^4 e^4 \mathrm{ln}\Lambda}
\end{equation}
where, $\epsilon$ is the inverse aspect ratio evaluated at the mid-radius, $(a/2)/R$, the ion temperature, $T_i$ is evaluated in joules, and the usual tokamak safety factor, $q$ has been replaced by the stellarator equivalent for a QHS stellarator, $1/\left(N - \stkout{\iota}\right)$ with $N$ the number of field periods.  The 0-D analysis indicates that for relatively modest amounts of external heating, small $\nu_i^*$ regimes can be realized.

The normalized pressure, $\beta$ is
\begin{equation}
    \beta = \frac{nT}{B^2/\left(2\mu_0\right)}
\end{equation}
with $\mu_0$ the permeability of free space.

The midscale design employs water cooled copper coils, and thus pulse length is expected to be limited by coil heating.

The parameters for the coil quality of fit and relevant engineering parameters for the midscale realization are given in table \ref{tab:coils}. These data include a preliminary analysis of the coils including a finite build made up of multiple filaments and a winding pack size commensurate with realizable current densities. The error $f_B$ is given by
\begin{equation}
    f_B \equiv \left[ \frac{1}{2} \int_{S} \left( \frac{\boldsymbol{B} \cdot \boldsymbol{n}}{\lvert \boldsymbol{B} \rvert} \right)^2 ds \right] \left( \int_S ds \right)^{-1},
    \label{eq:bncostfunction}
\end{equation}
where $S$ represents the boundary surface, and $\boldsymbol{B} \cdot \boldsymbol{n}$ represents the normal field on the boundary. More information on the procedure for generating these coils can be found in \citep{singh2020Optimization}.

\begin{table}
\begin{center}
\def~{\hphantom{0}}
\begin{tabular}{lcc}
     \hline
     \textbf{Param.} & \textbf{Single-filament} & \textbf{Multi-filament}  \\
     \hline
     Minimum Coil-Plasma Distance & 19.5 cm & 13.0 cm \\
     Average Coil-Plasma Distance & 22.5 cm & 16.0 cm \\
     Integrated Normal Field $f_B$ & {$0.147\mathrm{e}{-5}$} & {$0.143\mathrm{e}{-5}$} \\
     \hline
\end{tabular}
\end{center}
\caption{Parameters for the coil set produced for the WISTELL-A configuration using the dimensions in table \ref{tab:0d}}
\label{tab:coils}
\end{table}

\bibliographystyle{jpp}

\bibliography{bader_newQHS_2020}

\begin{thebibliography}{58}
\expandafter\ifx\csname natexlab\endcsname\relax\def\natexlab#1{#1}\fi
\def\au#1{#1} \def\ed#1{#1} \def\yr#1{#1}\def\at#1{#1}\def\jt#1{\textit{#1}}
  \def\bt#1{#1}\def\bvol#1{\textbf{#1}} \def\vol#1{#1} \def\pg#1{#1}
  \def\publ#1{#1}\def\arxiv#1{#1}\def\org#1{#1}\def\st#1{\textit{#1}}

\bibitem[Anderson {\em et~al.\/}(1995)Anderson, Almagri, Anderson, Matthews,
  Talmadge \& Shohet]{anderson1995helically}
{\sc \au{Anderson, F Simon~B}, \au{Almagri, Abdulgader~F}, \au{Anderson,
  David~T}, \au{Matthews, Peter~G}, \au{Talmadge, Joseph~N} \& \au{Shohet,
  J~Leon}} \yr{1995}  \at{The helically symmetric experiment,(hsx) goals,
  design and status}.  \jt{Fusion Technology}  \bvol{27}~(3T),  \pg{273--277}.

\bibitem[Bader {\em et~al.\/}(2017)Bader, Boozer, Hegna, Lazerson \&
  Schmitt]{bader2017hsx}
{\sc \au{Bader, A}, \au{Boozer, AH}, \au{Hegna, CC}, \au{Lazerson, SA} \&
  \au{Schmitt, JC}} \yr{2017}  \at{Hsx as an example of a resilient
  non-resonant divertor}.  \jt{Physics of Plasmas}  \bvol{24}~(3),
  \pg{032506}.

\bibitem[Bader {\em et~al.\/}(2019)Bader, Drevlak, Anderson, Faber, Hegna,
  Likin, Schmitt \& Talmadge]{bader2019stellarator}
{\sc \au{Bader, Aaron}, \au{Drevlak, M}, \au{Anderson, DT}, \au{Faber, BJ},
  \au{Hegna, CC}, \au{Likin, KM}, \au{Schmitt, JC} \& \au{Talmadge, JN}}
  \yr{2019}  \at{Stellarator equilibria with reactor relevant energetic
  particle losses}.  \jt{Journal of Plasma Physics}  \bvol{85}~(5).

\bibitem[Bader {\em et~al.\/}(2018)Bader, Hegna, Cianciosa \&
  Hartwell]{bader2018minimum}
{\sc \au{Bader, Aaron}, \au{Hegna, CC}, \au{Cianciosa, M} \& \au{Hartwell, GJ}}
  \yr{2018}  \at{Minimum magnetic curvature for resilient divertors using
  compact toroidal hybrid geometry}.  \jt{Plasma Physics and Controlled Fusion}
   \bvol{60}~(5),  \pg{054003}.

\bibitem[Bauer {\em et~al.\/}(1984)Bauer, Betancourt \&
  Garabedian]{bauer1984mhd}
{\sc \au{Bauer, F.}, \au{Betancourt, 0.} \& \au{Garabedian, P.A.}} \yr{1984}
  {\em Magnetohydrodynamic Equilibrium and Stability of Stellarators\/}.
  \publ{Springer-Verlag, New York}.

\bibitem[Beidler {\em et~al.\/}(1990)Beidler, Grieger, Herrnegger, Harmeyer,
  Kisslinger, Lotz, Maassberg, Merkel, N{\"u}hrenberg, Rau {\em
  et~al.\/}]{beidler1990physics}
{\sc \au{Beidler, Craig}, \au{Grieger, G{\"u}nter}, \au{Herrnegger, Franz},
  \au{Harmeyer, Ewald}, \au{Kisslinger, Johann}, \au{Lotz, Wolf},
  \au{Maassberg, Henning}, \au{Merkel, Peter}, \au{N{\"u}hrenberg, J{\"u}rgen},
  \au{Rau, Fritz} \& \au{others}} \yr{1990}  \at{Physics and engineering design
  for wendelstein vii-x}.  \jt{Fusion Technology}  \bvol{17}~(1),
  \pg{148--168}.

\bibitem[Beidler {\em et~al.\/}(2001)Beidler, Harmeyer, Herrnegger, Igitkhanov,
  Kendl, Kisslinger, Kolesnichenko, Lutsenko, N{\"u}hrenberg, Sidorenko {\em
  et~al.\/}]{beidler2001helias}
{\sc \au{Beidler, CD}, \au{Harmeyer, E}, \au{Herrnegger, F}, \au{Igitkhanov,
  Yu}, \au{Kendl, A}, \au{Kisslinger, J}, \au{Kolesnichenko, Ya~I},
  \au{Lutsenko, VV}, \au{N{\"u}hrenberg, C}, \au{Sidorenko, I} \& \au{others}}
  \yr{2001}  \at{The helias reactor hsr4/18}.  \jt{Nuclear Fusion}
  \bvol{41}~(12),  \pg{1759}.

\bibitem[Boozer(1981)]{boozer1981coordinates}
{\sc \au{Boozer, A~H}} \yr{1981}  \at{Plasma equilibrium with rational magnetic
  surfaces}.  \jt{Physics of Fluids}  \bvol{24}~(11),  \pg{1999}.

\bibitem[Boozer \& Gardner(1990)]{boozer1990BOOTSJ}
{\sc \au{Boozer, A~H} \& \au{Gardner, H.~J.}} \yr{1990}  \at{The bootstrap
  current in stellarators}.  \jt{Physics of Fluids B-Plasma Physics}
  \bvol{2}~(10),  \pg{2408}.

\bibitem[Boozer \& Punjabi(2018)]{boozer2018simulation}
{\sc \au{Boozer, Allen~H} \& \au{Punjabi, Alkesh}} \yr{2018}  \at{Simulation of
  stellarator divertors}.  \jt{Physics of Plasmas}  \bvol{25}~(9),
  \pg{092505}.

\bibitem[Brent(2013)]{brent2013algorithms}
{\sc \au{Brent, Richard~P}} \yr{2013} {\em Algorithms for minimization without
  derivatives\/}.  \publ{Courier Corporation}.

\bibitem[Canik {\em et~al.\/}(2007)Canik, Anderson, Anderson, Likin, Talmadge
  \& Zhai]{canik2007experiment}
{\sc \au{Canik, J.~M.}, \au{Anderson, D.~T.}, \au{Anderson, F. S.~B.},
  \au{Likin, K.~M.}, \au{Talmadge, J.~N.} \& \au{Zhai, K.}} \yr{2007}
  \at{Experimental demonstration of improved neoclassical transport with
  quasihelical symmetry}.  \jt{Phys. Rev. Lett.}  \bvol{98},  \pg{085002}.

\bibitem[Carreras {\em et~al.\/}(1988)Carreras, Dominguez, Garcia, Lynch, Lyon,
  Cary, Hanson \& Navarro]{carreras1998torsatronconfigs}
{\sc \au{Carreras, B.A.}, \au{Dominguez, N.}, \au{Garcia, L.}, \au{Lynch,
  V.E.}, \au{Lyon, J.F.}, \au{Cary, J.R.}, \au{Hanson, J.D.} \& \au{Navarro,
  A.P.}} \yr{1988}  \at{Low-aspect-ratio torsatron configurations}.
  \jt{Nuclear Fusion}  \bvol{28},  \pg{1195--1207}.

\bibitem[Drevlak {\em et~al.\/}(2018)Drevlak, Beidler, Geiger, Helander \&
  Turkin]{drevlak2018optimisation}
{\sc \au{Drevlak, M}, \au{Beidler, CD}, \au{Geiger, J}, \au{Helander, P} \&
  \au{Turkin, Y}} \yr{2018}  \at{Optimisation of stellarator equilibria with
  rose}.  \jt{Nuclear Fusion}  \bvol{59}~(1),  \pg{016010}.

\bibitem[Drevlak {\em et~al.\/}(2014)Drevlak, Geiger, Helander \&
  Turkin]{drevlak2014fast}
{\sc \au{Drevlak, M}, \au{Geiger, J}, \au{Helander, P} \& \au{Turkin, Yu}}
  \yr{2014}  \at{Fast particle confinement with optimized coil currents in the
  w7-x stellarator}.  \jt{Nuclear Fusion}  \bvol{54}~(7),  \pg{073002}.

\bibitem[Faber {\em et~al.\/}(2018)Faber, Pueschel, Terry, Hegna \&
  Roman]{faber2018jpp}
{\sc \au{Faber, B. J.}, \au{Pueschel, M. J.}, \au{Terry, P. W.}, \au{Hegna,
  C. C.} \& \au{Roman, J. E.}} \yr{2018}  \at{Stellarator microinstabilities
  and turbulence at low magnetic shear}.  \jt{Journal of Plasma Physics}
  \bvol{84}~(5),  \pg{905840503}.

\bibitem[Faber(2020)]{faber2020ptsm3d}
{\sc \au{Faber, B.~J.}} \yr{2020} {PTSM3D}.
  {h}ttps://doi.org/10.5281/zenodo.3726923.

\bibitem[Faber {\em et~al.\/}(2015)Faber, Pueschel, Proll, Xanthopoulos, Terry,
  Hegna, Weir, Likin \& Talmadge]{faber2015pop}
{\sc \au{Faber, B.~J.}, \au{Pueschel, M.~J.}, \au{Proll, J. H.~E.},
  \au{Xanthopoulos, P.}, \au{Terry, P.~W.}, \au{Hegna, C.~C.}, \au{Weir,
  G.~M.}, \au{Likin, K.~M.} \& \au{Talmadge, J.~N.}} \yr{2015}  \at{Gyrokinetic
  studies of trapped electron mode turbulence in the {H}elically {S}ymmetric
  e{X}periment stellarator}.  \jt{Physics of Plasmas}  \bvol{22},  \pg{072305}.

\bibitem[Feng {\em et~al.\/}(2004)Feng, Sardei, Kisslinger, Grigull, McCormick
  \& Reiter]{feng20043d}
{\sc \au{Feng, Y}, \au{Sardei, F}, \au{Kisslinger, J}, \au{Grigull, P},
  \au{McCormick, K} \& \au{Reiter, D}} \yr{2004}  \at{3d edge modeling and
  island divertor physics}.  \jt{Contributions to Plasma Physics}
  \bvol{44}~(1-3),  \pg{57--69}.

\bibitem[Greene(1997)]{greene1997brief}
{\sc \au{Greene, John~M}} \yr{1997}  \at{A brief review of magnetic wells}.
  \jt{Comments on Plasma Physics and Controlled Fusion}  \bvol{17},
  \pg{389--402}.

\bibitem[Grieger {\em et~al.\/}(1992)Grieger, Lotz, Merkel, N{\"u}hrenberg,
  Sapper, Strumberger, Wobig, Burhenn, Erckmann, Gasparino {\em
  et~al.\/}]{grieger1992physics}
{\sc \au{Grieger, G}, \au{Lotz, W}, \au{Merkel, P}, \au{N{\"u}hrenberg, J},
  \au{Sapper, J}, \au{Strumberger, E}, \au{Wobig, H}, \au{Burhenn, R},
  \au{Erckmann, V}, \au{Gasparino, U} \& \au{others}} \yr{1992}  \at{Physics
  optimization of stellarators}.  \jt{Physics of Fluids B: Plasma Physics}
  \bvol{4}~(7),  \pg{2081--2091}.

\bibitem[Hegna \& Callen(1994)]{hegna1994islands}
{\sc \au{Hegna, C.C.} \& \au{Callen, J.~D.}} \yr{1994}  \at{Stability of
  boostrap current driven magnetic islands in stellarators}.  \jt{Phys.
  Plasmas}  \bvol{1}~(9),  \pg{3135}.

\bibitem[Hegna \& Hudson(2001)]{hegna2001ballooning}
{\sc \au{Hegna, C.C.} \& \au{Hudson, S.~R.}} \yr{2001}  \at{Loss of second
  ballooning stability in three-dimensional equilibria}.  \jt{Phys. Rev. Lett.}
   \bvol{87}~(3),  \pg{035001}.

\bibitem[Hegna {\em et~al.\/}(2018)Hegna, Terry \& Faber]{hegna2018theory}
{\sc \au{Hegna, C.C.}, \au{Terry, P.W.} \& \au{Faber, B.J.}} \yr{2018}
  \at{Theory of itg turbulent saturation in stellarators: identifying
  mechanisms to reduce turbulent transport}.  \jt{Physics of Plasmas}
  \bvol{25}~(2),  \pg{022511}.

\bibitem[Henneberg {\em et~al.\/}(2019{\natexlab{{\em a\/}}})Henneberg, Drevlak
  \& Helander]{henneberg2019improving}
{\sc \au{Henneberg, SA}, \au{Drevlak, M} \& \au{Helander, P}}
  \yr{2019{\natexlab{{\em a\/}}}}  \at{Improving fast-particle confinement in
  quasi-axisymmetric stellarator optimization}.  \jt{Plasma Physics and
  Controlled Fusion}  \bvol{62}~(1),  \pg{014023}.

\bibitem[Henneberg {\em et~al.\/}(2019{\natexlab{{\em b\/}}})Henneberg,
  Drevlak, N{\"u}hrenberg, Beidler, Turkin, Loizu \&
  Helander]{henneberg2019properties}
{\sc \au{Henneberg, SA}, \au{Drevlak, M}, \au{N{\"u}hrenberg, C}, \au{Beidler,
  CD}, \au{Turkin, Y}, \au{Loizu, J} \& \au{Helander, P}}
  \yr{2019{\natexlab{{\em b\/}}}}  \at{Properties of a new quasi-axisymmetric
  configuration}.  \jt{Nuclear Fusion}  \bvol{59}~(2),  \pg{026014}.

\bibitem[Hirshman \& Whitson(1983)]{Hirshman_PoF_1983}
{\sc \au{Hirshman, Steven~P} \& \au{Whitson, JC}} \yr{1983}
  \at{Steepest-descent moment method for three-dimensional magnetohydrodynamic
  equilibria}.  \jt{The Physics of fluids}  \bvol{26}~(12),  \pg{3553--3568}.

\bibitem[Jenko {\em et~al.\/}(2000)Jenko, Dorland, Kotschenreuther \&
  Rogers]{jenko2000gene}
{\sc \au{Jenko, F.}, \au{Dorland, W.}, \au{Kotschenreuther, M.} \& \au{Rogers,
  B.~N.}} \yr{2000}  \at{Electron temperature gradient driven turbulence}.
  \jt{Physics of Plasmas}  \bvol{7}~(5),  \pg{1904--1910}.

\bibitem[K{\"o}nig {\em et~al.\/}(2002)K{\"o}nig, Grigull, McCormick, Feng,
  Kisslinger, Komori, Masuzaki, Matsuoka, Obiki, Ohyabu {\em
  et~al.\/}]{konig2002divertor}
{\sc \au{K{\"o}nig, R}, \au{Grigull, P}, \au{McCormick, K}, \au{Feng, Y},
  \au{Kisslinger, J}, \au{Komori, A}, \au{Masuzaki, Suguru}, \au{Matsuoka, K},
  \au{Obiki, T}, \au{Ohyabu, N} \& \au{others}} \yr{2002}  \at{The divertor
  program in stellarators}.  \jt{Plasma physics and controlled fusion}
  \bvol{44}~(11),  \pg{2365}.

\bibitem[Ku {\em et~al.\/}(2008)Ku, Garabedian, Lyon, Turnbull, Grossman, Mau,
  Zarnstorff \& Team]{ku2008physics}
{\sc \au{Ku, LP}, \au{Garabedian, PR}, \au{Lyon, J}, \au{Turnbull, A},
  \au{Grossman, A}, \au{Mau, TK}, \au{Zarnstorff, M} \& \au{Team, ARIES}}
  \yr{2008}  \at{Physics design for aries-cs}.  \jt{Fusion Science and
  Technology}  \bvol{54}~(3),  \pg{673--693}.

\bibitem[Landreman(2017)]{landreman2017improved}
{\sc \au{Landreman, Matt}} \yr{2017}  \at{An improved current potential method
  for fast computation of stellarator coil shapes}.  \jt{Nuclear Fusion}
  \bvol{57}~(4),  \pg{046003}.

\bibitem[Landreman \& Paul(2018)]{landreman2018computing}
{\sc \au{Landreman, Matt} \& \au{Paul, Elizabeth}} \yr{2018}  \at{Computing
  local sensitivity and tolerances for stellarator physics properties using
  shape gradients}.  \jt{Nuclear Fusion}  \bvol{58}~(7),  \pg{076023}.

\bibitem[Landreman \& Sengupta(2018)]{landreman2018direct}
{\sc \au{Landreman, Matt} \& \au{Sengupta, Wrick}} \yr{2018}  \at{Direct
  construction of optimized stellarator shapes. part 1. theory in cylindrical
  coordinates}.  \jt{Journal of Plasma Physics}  \bvol{84}~(6).

\bibitem[Landreman {\em et~al.\/}(2019)Landreman, Sengupta \&
  Plunk]{landreman2019direct}
{\sc \au{Landreman, Matt}, \au{Sengupta, Wrick} \& \au{Plunk, Gabriel~G}}
  \yr{2019}  \at{Direct construction of optimized stellarator shapes. part 2.
  numerical quasisymmetric solutions}.  \jt{Journal of Plasma Physics}
  \bvol{85}~(1).

\bibitem[Landreman {\em et~al.\/}(2014)Landreman, Smith, Mollén \&
  Helander]{landreman2014sfincs}
{\sc \au{Landreman, M.}, \au{Smith, H.M.}, \au{Mollén, A.} \& \au{Helander,
  P.}} \yr{2014}  \at{Comparison of particle trajectories and collision
  operators for collisional transport in nonaxisymmetric plasmas}.  \jt{Physics
  of Plasmas}  \bvol{21},  \pg{042503}.

\bibitem[McKinney {\em et~al.\/}(2019)McKinney, Pueschel, Faber, Hegna,
  Talmadge, Anderson, Mynick \& Xanthopoulos]{mckinney2019comparison}
{\sc \au{McKinney, IJ}, \au{Pueschel, MJ}, \au{Faber, BJ}, \au{Hegna, CC},
  \au{Talmadge, JN}, \au{Anderson, DT}, \au{Mynick, HE} \& \au{Xanthopoulos,
  P}} \yr{2019}  \at{A comparison of turbulent transport in a quasi-helical and
  a quasi-axisymmetric stellarator}.  \jt{Journal of Plasma Physics}
  \bvol{85}~(5).

\bibitem[Miller {\em et~al.\/}(1996)Miller, Team {\em
  et~al.\/}]{miller1996stellarator}
{\sc \au{Miller, R}, \au{Team, SPPS} \& \au{others}} \yr{1996}  \at{The
  stellarator power plant study}.  \jt{University of California San Diego
  Report UCSD-ENG-004} .

\bibitem[Mynick {\em et~al.\/}(2010)Mynick, Pomphrey \&
  Xanthopoulos]{mynick2010optimizing}
{\sc \au{Mynick, HE}, \au{Pomphrey, N} \& \au{Xanthopoulos, P}} \yr{2010}
  \at{Optimizing stellarators for turbulent transport}.  \jt{Physical review
  letters}  \bvol{105}~(9),  \pg{095004}.

\bibitem[Nemov {\em et~al.\/}(1999)Nemov, Kasilov, Kernbichler \&
  Heyn]{nemov1999evaluation}
{\sc \au{Nemov, VV}, \au{Kasilov, SV}, \au{Kernbichler, W} \& \au{Heyn, MF}}
  \yr{1999}  \at{Evaluation of 1/$\nu$ neoclassical transport in stellarators}.
   \jt{Physics of plasmas}  \bvol{6}~(12),  \pg{4622--4632}.

\bibitem[Nemov {\em et~al.\/}(2008)Nemov, Kasilov, Kernbichler \&
  Leitold]{nemov2008poloidal}
{\sc \au{Nemov, VV}, \au{Kasilov, SV}, \au{Kernbichler, Winfried} \&
  \au{Leitold, GO}} \yr{2008}  \at{Poloidal motion of trapped particle orbits
  in real-space coordinates}.  \jt{Physics of plasmas}  \bvol{15}~(5),
  \pg{052501}.

\bibitem[Nührenberg \& Zille(1988)]{nuhrenberg1988qhs}
{\sc \au{Nührenberg, J.} \& \au{Zille, R.}} \yr{1988}  \at{Quasi-helically
  symmetric toroidal stellarators}.  \jt{Physics Letters A}  \bvol{129}~(2),
  \pg{113--117}.

\bibitem[Pablant {\em et~al.\/}(2020)]{pablant2020experiment}
{\sc \au{Pablant, N.} \& \au{others}} \yr{2020}  \at{Investigation of the
  neoclassical ambipolar electric field in ion-root plasmas on {W7-X}}.
  \jt{Nuclear Fusion}  \bvol{60}~(3),  \pg{036021}.

\bibitem[Plunk {\em et~al.\/}(2019)Plunk, Landreman \&
  Helander]{plunk2019direct}
{\sc \au{Plunk, Gabriel~G}, \au{Landreman, Matt} \& \au{Helander, Per}}
  \yr{2019}  \at{Direct construction of optimized stellarator shapes. iii.
  omnigenity near the magnetic axis}.  \jt{arXiv preprint arXiv:1909.08919} .

\bibitem[Plunk {\em et~al.\/}(2017)Plunk, Xanthopoulos \&
  Helander]{plunk2017saturation}
{\sc \au{Plunk, G.~G.}, \au{Xanthopoulos, P.} \& \au{Helander, P.}} \yr{2017}
  \at{Distinct turbulence saturation regimes in stellarators}.  \jt{Phys. Rev.
  Lett.}  \bvol{118},  \pg{105002}.

\bibitem[Pueschel {\em et~al.\/}(2016)Pueschel, Faber, Citrin, Hegna, Terry \&
  Hatch]{pueschel2016prl}
{\sc \au{Pueschel, M.~J.}, \au{Faber, B.~J.}, \au{Citrin, J.}, \au{Hegna,
  C.~C.}, \au{Terry, P.~W.} \& \au{Hatch, D.~R.}} \yr{2016}  \at{Stellarator
  turbulence: Subdominant eigenmodes and quasilinear modeling}.  \jt{Phys. Rev.
  Lett.}  \bvol{116},  \pg{085001}.

\bibitem[Sanchez {\em et~al.\/}(2000)Sanchez, Hirshman \&
  Ware]{sanchez2000cobra}
{\sc \au{Sanchez, R.}, \au{Hirshman, S.~P.} \& \au{Ware, A.~S.}} \yr{2000}
  \at{Cobra:an optimized code for fast evaluation of ideal ballooning stability
  of three dimensional equilibria}.  \jt{J. Comp. Phys.}  \bvol{161}~(2),
  \pg{576}.

\bibitem[Schmitt {\em et~al.\/}(2013)Schmitt, Talmadge \&
  Anderson]{schmitt2013ps}
{\sc \au{Schmitt, J.C.}, \au{Talmadge, J.N.} \& \au{Anderson, D.T.}} \yr{2013}
  \at{Measurement of a helical pfirsch-schlüuter current with reduced
  magnitude in hsx}.  \jt{Nuclear Fusion}  \bvol{53},  \pg{0820001}.

\bibitem[Schmitt {\em et~al.\/}(2014)Schmitt, Talmadge, Anderson \&
  Hanson]{schmitt2014bs}
{\sc \au{Schmitt, J.C.}, \au{Talmadge, J.N.}, \au{Anderson, D.T.} \&
  \au{Hanson, J.D.}} \yr{2014}  \at{Modeling, measurement and 3-d equilibrium
  reconstruction of the bootstrap current in the helically symmetric
  experiment}.  \jt{Physics of Plasmas}  \bvol{21},  \pg{092518}.

\bibitem[Shimizu {\em et~al.\/}(2018)Shimizu, Liu, Isobe, Okamura, Nishimura,
  Suzuki, Xu, Zhang, Liu, Huang {\em et~al.\/}]{shimizu2018configuration}
{\sc \au{Shimizu, Akihiro}, \au{Liu, Haifeng}, \au{Isobe, Mitsutaka},
  \au{Okamura, Shoichi}, \au{Nishimura, Shin}, \au{Suzuki, Chihiro}, \au{Xu,
  Yuhong}, \au{Zhang, Xin}, \au{Liu, Bing}, \au{Huang, Jie} \& \au{others}}
  \yr{2018}  \at{Configuration property of the chinese first quasi-axisymmetric
  stellarator}.  \jt{Plasma and Fusion Research}  \bvol{13},
  \pg{3403123--3403123}.

\bibitem[Singh {\em et~al.\/}(2020)Singh, Kruger, Bader, Zhu, Hudson \&
  Anderston]{singh2020Optimization}
{\sc \au{Singh, L.}, \au{Kruger, T.G.}, \au{Bader, A.}, \au{Zhu, C.},
  \au{Hudson, S.R.} \& \au{Anderston, D.T.}} \yr{2020}  \at{Optimization of
  finite-build stellarator coils}.  \jt{Submitted to Journal of Plasma Physics}
  .

\bibitem[Strumberger(1992)]{strumberger1992magnetic}
{\sc \au{Strumberger, Erika}} \yr{1992}  \at{Magnetic field line diversion in
  helias stellarator configurations: perspectives for divertor operation}.
  \jt{Nuclear fusion}  \bvol{32}~(5),  \pg{737}.

\bibitem[Sudo {\em et~al.\/}(1990)Sudo, Takeiri, Zushi, Sano, Itoh, Kondo \&
  Iiyoshi]{sudo1990scalings}
{\sc \au{Sudo, S}, \au{Takeiri, Y}, \au{Zushi, H}, \au{Sano, F}, \au{Itoh, K},
  \au{Kondo, K} \& \au{Iiyoshi, A}} \yr{1990}  \at{Scalings of energy
  confinement and density limit in stellarator/heliotron devices}.  \jt{Nuclear
  Fusion}  \bvol{30}~(1),  \pg{11}.

\bibitem[Talmadge {\em et~al.\/}(2001)Talmadge, Sakaguchi, Anderson, Anderson
  \& Almagri]{talmadge2001orbits}
{\sc \au{Talmadge, J.N.}, \au{Sakaguchi, V.}, \au{Anderson, F. S.~B.},
  \au{Anderson, D.~T.} \& \au{Almagri, A.~F.}} \yr{2001}  \at{Experimental
  determination of the magnetic field spectrum in the helically symmetric
  experiment using passing particle orbits}.  \jt{Physics of Plasmas}
  \bvol{8},  \pg{5165}.

\bibitem[Weller {\em et~al.\/}(2001)Weller, Anton, Geiger, Hirsch, Jaenicke,
  Werner, Team, N\"uhrenberg, Sallander \& Spong]{weller2001MHDstability}
{\sc \au{Weller, A.}, \au{Anton, M.}, \au{Geiger, J.}, \au{Hirsch, R.},
  \au{Jaenicke, R.}, \au{Werner, A.}, \au{Team, W7-AS}, \au{N\"uhrenberg, J.},
  \au{Sallander, E.} \& \au{Spong, D.~A.}} \yr{2001}  \at{Survey of
  magnetohydrodynamic instabilities in the advanced stellarator wendelstein
  7-as}.  \jt{Phys. Plasmas}  \bvol{8}~(3),  \pg{931}.

\bibitem[Xanthopoulos {\em et~al.\/}(2014)Xanthopoulos, Mynick, Helander,
  Turkin, Plunk, Jenko, G{\"o}rler, Told, Bird \&
  Proll]{xanthopoulos2014controlling}
{\sc \au{Xanthopoulos, P}, \au{Mynick, HE}, \au{Helander, P}, \au{Turkin, Yu},
  \au{Plunk, GG}, \au{Jenko, F}, \au{G{\"o}rler, T}, \au{Told, D}, \au{Bird, T}
  \& \au{Proll, JHE}} \yr{2014}  \at{Controlling turbulence in present and
  future stellarators}.  \jt{Physical review letters}  \bvol{113}~(15),
  \pg{155001}.

\bibitem[Yamada {\em et~al.\/}(2005)Yamada, Harris, Dinklage, Ascasibar, Sano,
  Okamura, Talmadge, Stroth, Kus, Murakami {\em
  et~al.\/}]{yamada2005characterization}
{\sc \au{Yamada, H}, \au{Harris, JH}, \au{Dinklage, Andreas}, \au{Ascasibar,
  E}, \au{Sano, F}, \au{Okamura, S}, \au{Talmadge, J}, \au{Stroth, U}, \au{Kus,
  A}, \au{Murakami, S} \& \au{others}} \yr{2005}  \at{Characterization of
  energy confinement in net-current free plasmas using the extended
  international stellarator database}.  \jt{Nuclear Fusion}  \bvol{45}~(12),
  \pg{1684}.

\bibitem[Zarnstorff {\em et~al.\/}(2001)Zarnstorff, Berry, Brooks, Fredrickson,
  Fu, Hirshman, Hudson, Ku, Lazarus, Mikkelsen {\em
  et~al.\/}]{zarnstorff2001physics}
{\sc \au{Zarnstorff, MC}, \au{Berry, LA}, \au{Brooks, A}, \au{Fredrickson, E},
  \au{Fu, GY}, \au{Hirshman, S}, \au{Hudson, S}, \au{Ku, LP}, \au{Lazarus, E},
  \au{Mikkelsen, D} \& \au{others}} \yr{2001}  \at{Physics of the compact
  advanced stellarator ncsx}.  \jt{Plasma Physics and Controlled Fusion}
  \bvol{43}~(12A),  \pg{A237}.

\bibitem[Zhu {\em et~al.\/}(2018)Zhu, Hudson, Song \& Wan]{zhu2018designing}
{\sc \au{Zhu, Caoxiang}, \au{Hudson, Stuart~R}, \au{Song, Yuntao} \& \au{Wan,
  Yuanxi}} \yr{2018}  \at{Designing stellarator coils by a modified newton
  method using focus}.  \jt{Plasma Physics and Controlled Fusion}
  \bvol{60}~(6),  \pg{065008}.

\end{thebibliography}

\end{document}